\documentclass[]{antgroup}

\PassOptionsToPackage{numbers, compress, sort}{natbib}
\usepackage{antgroup}
\usepackage[utf8]{inputenc} 
\usepackage[T1]{fontenc}    
\usepackage{hyperref}       
\usepackage{url}            
\usepackage{booktabs}       
\usepackage{amsfonts}       
\usepackage{nicefrac}       
\usepackage{microtype}      
\usepackage{xcolor}         
\usepackage{xspace}
\usepackage{colortbl}
\usepackage[most]{tcolorbox}
\usepackage{tabularx}
\usepackage{fontawesome5}

\newcommand{\guard}{{SingProbe}\xspace}

\newcommand{\partitle}[1]{\noindent \textbf{#1.}}

\newcommand{\githubicon}{%
    \raisebox{-0.15em}{%
        \includegraphics[height=1.1em]{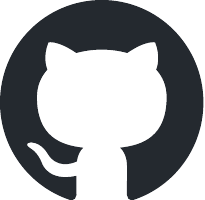}%
    }%
}

\newcommand{\hficon}{%
    \raisebox{-0.15em}{%
        \includegraphics[height=1.1em]{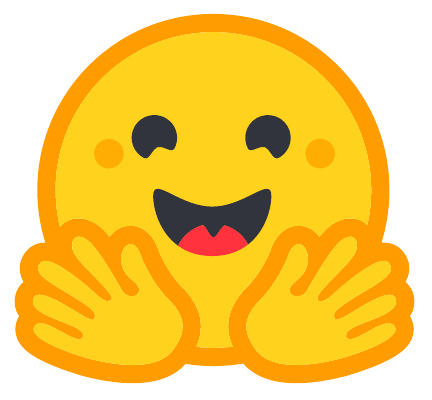}%
    }%
}

\title{\guard Technical Report}

\author{
  Sing Team \\
  AI Security Lab, Ant Group
}

\begin{document}

\maketitle

\begin{center}
\small
\href{https://github.com/inclusionAI/SingProbe}{\githubicon\ https://github.com/inclusionAI/SingProbe}
\\
\href{https://huggingface.co/collections/inclusionAI/singprobe}{\hficon\ https://huggingface.co/collections/inclusionAI/singprobe}
\end{center}

\begin{figure}[h!]
    \centering
    \includegraphics[width=0.85\linewidth]{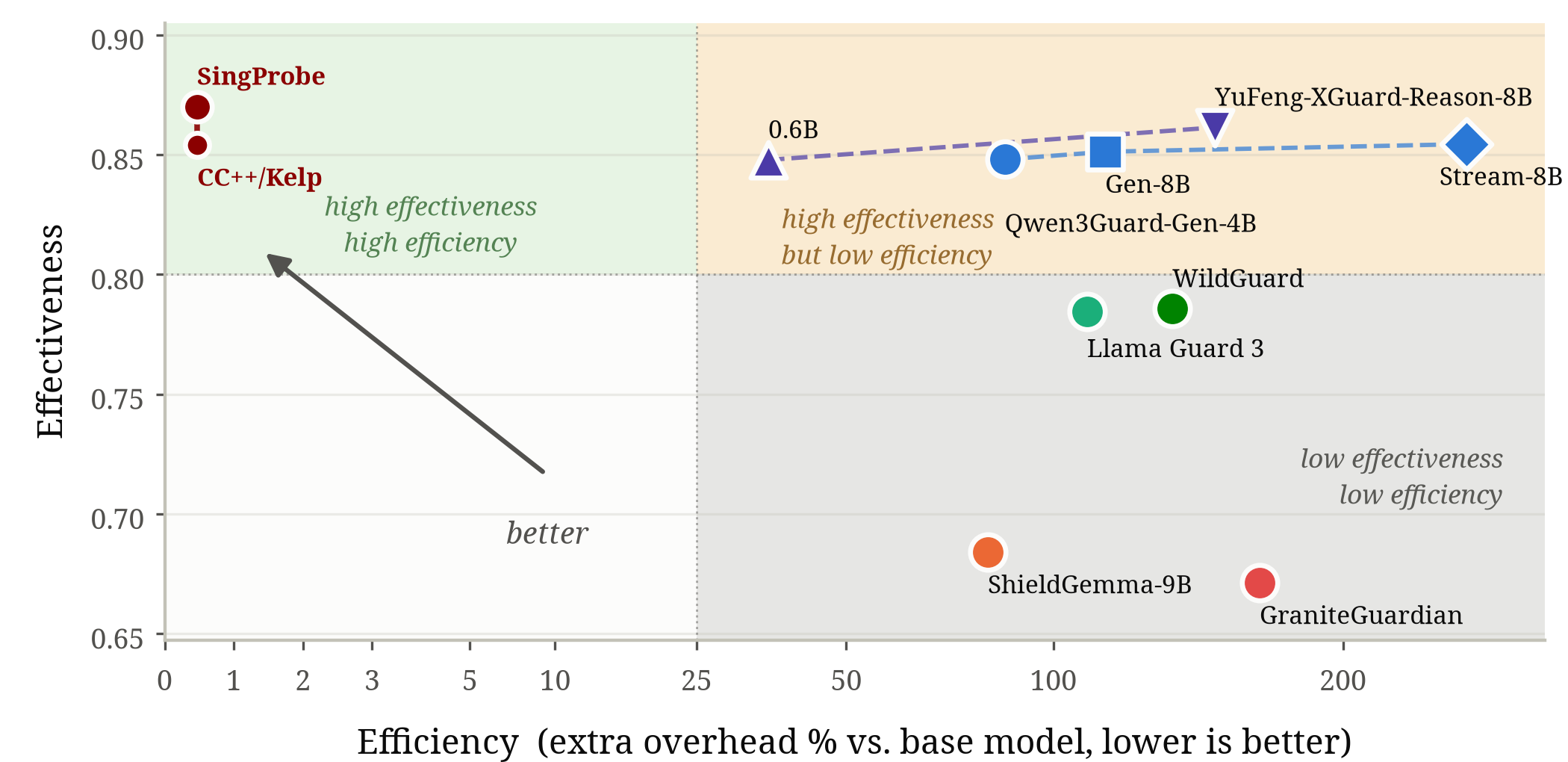}
     \caption{Effectiveness–efficiency trade-off of guardrail models. Efficiency (x) is the extra inference latency of a guard over the base-model-only serving baseline. We use Ling-3.0-flash as the base model. A non-streaming guard is invoked once after the complete response has been generated, whereas a streaming guard is invoked at every generation step. Effectiveness (y-axis) is measured as the average of the guard's Query F1 and Response F1 across the evaluated benchmarks. CC++/Kelp are our reproduced versions of two intrinsic guardrails~\citep{cunningham2026CC++, li2025kelp}, using the same model architecture and training data, with differences only in the training methodology.}
    \label{fig:tradeoff}
    \vspace{1em}
\end{figure}

\begin{abstract}
  We present \guard, an open intrinsic guardrail framework for generation-time monitoring of large language models (LLMs). Intrinsic guardrails reuse hidden states already produced by the base model during autoregressive decoding, rather than relying on an independent model to repeatedly process generated text. While this route has been explored in industrial systems, the community lacks a broadly reusable open stack that combines cross-model guard adaptations, unified training methods, serving integrations, and systematic evaluation resources. SingProbe is designed to provide this missing layer.
  SingProbe uses a lightweight probe to continuously produce query-intent, response-safety, and hallucination-risk signals during decoding. These signals support both response-level decisions and token-level risk localization. This report describes the full intrinsic-guardrail stack: training methods, serving integrations with SGLang and vLLM, and adapted guard models for 29 open-source base models across diverse families and scales. We also introduce SingStreamBench, a benchmark that measures whether streaming guardrails remain inactive on benign prefixes while promptly detecting emerging unsafe content.
  Across evaluations of safety, streaming detection, hallucination detection, false-positive robustness, online monitoring, and runtime overhead, SingProbe provides performance competitive with, and in several settings stronger than, state-of-the-art standalone guardrails and specialized hallucination detectors, while adding less than 0.5\% serving overhead in our implementation. Beyond passive monitoring, we show that intrinsic guard signals can guide constrained decoding and selectively activate medical-risk interventions in SingProbe-Med. By open-sourcing our infrastructure, training methods, and model adaptations, we aim to facilitate the broader adoption and deployment of intrinsic guardrails, as well as further research in this direction.

\end{abstract}

\section{Introduction}

In this technical report, we propose \guard (\textbf{\underline{S}afety \underline{in} \underline{G}eneration \underline{Probe}}), an open intrinsic guardrail framework for generation-time monitoring of large language models (LLMs). This report describes the complete SingProbe stack: its training methods, adapted guard models, serving integrations, and evaluation resources. Our objective is to make intrinsic guardrails a practical and deployable component, so that model developers and serving-system builders can train, deploy, and evaluate guardrails that operate within the generation pipeline.

SingProbe is an intrinsic guardrail that reuses the hidden states that a base model already produces during autoregressive decoding, and applies a lightweight prediction module to emit risk signals. This differs from an external guardrail, which deploys a separate model to process the input or generated text independently~\citep{zhao2025qwen3guard,inan2023llamaguard,lin2026yufengxguard}. The intrinsic route has already been explored in industrial systems, including Anthropic's Constitutional Classifiers++~\citep{cunningham2026CC++} and Alibaba Group's Kelp/PlugGuard~\citep{li2025kelp}. These efforts demonstrate the value of using internal representations for generation-time safety monitoring, but have primarily been developed for particular models or production settings. A broadly reusable open stack, including cross-model guard adaptations, unified training methods, standard serving integrations, and systematic evaluation resources, has not yet emerged. SingProbe is designed to provide this missing layer for the community.

SingProbe converts base-model hidden states into a unified stream of guard signals. During generation, its lightweight probe continuously predicts query intent, response safety, and response hallucination risk. The resulting signals support two complementary uses: response-level decisions over completed trajectories and token-level localization of risks as they emerge. The training method is designed to learn useful streaming signals from commonly available coarse response-level supervision, enabling the same framework to support multiple monitoring tasks without requiring a separate text encoder.

Importantly, we make this framework deployable in common LLM serving systems. SingProbe is integrated with SGLang\footnote{\url{https://github.com/jinzhen-lin/sglang/tree/token-probe-ling3-flash-main}}~\citep{zheng2024sglang} and vLLM\footnote{\url{https://github.com/jinzhen-lin/vllm/tree/bailing-v3-token-probe}}~\citep{kwon2023vllm}, so decoding and guard scoring run in the same serving pipeline. This avoids repeatedly encoding generated text with an additional guard model and removes the need to coordinate a separate guardrail service. To support adoption across model families, we adapt and release SingProbe for 29 open-source base models spanning Ling 3.0, Qwen, Llama 3, Gemma 4, gpt-oss, GLM, Step, MiniMax, DeepSeek, and Hy. Each adaptation exposes the same interface for query intent, response safety, and hallucination monitoring.

We evaluate this stack across query-intent classification, response safety, streaming detection, hallucination detection, false-positive robustness, online monitoring, and runtime overhead. We also introduce \textbf{SingStreamBench}, which evaluates a guardrail's temporal behavior: it should remain inactive on benign prefixes and respond promptly when unsafe content emerges. Across these evaluations, SingProbe substantially reduces additional serving overhead while achieving performance competitive with, and in several settings stronger than, state-of-the-art standalone guardrails and specialized hallucination detectors.

Finally, SingProbe treats guard outputs as runtime control signals rather than only as classifier outputs. We study online monitoring under free autoregressive generation and SingProbe-guided constrained decoding. We further introduce \textbf{SingProbe-Med}, which uses intrinsic signals to selectively activate medical-risk intervention on relevant suffixes instead of applying intervention continuously. These studies illustrate how the same monitoring interface can support downstream generation policies.

In summary, this report makes the following contributions:

\begin{enumerate}[leftmargin=*]
    \item We present SingProbe, an intrinsic guardrail framework that reuses LLM hidden states to continuously monitor query intent, response safety, and hallucination risk during generation.
    \item We provide an open intrinsic-guardrail stack comprising training methods, adapted models for 29 open-source base models, and serving integrations with SGLang and vLLM.
    \item We introduce SingStreamBench and systematically evaluate safety, hallucination, streaming behavior, false positives, online monitoring, and runtime overhead.
    \item We demonstrate that intrinsic guard signals can support generation-time control, including constrained decoding and selective medical-risk intervention.
\end{enumerate}

\section{Methodology}

\subsection{Formulation}

Let $\mathcal{M}$ denote an autoregressive LLM with $L$ layers. Given a token sequence $\mathbf{x}=(x_1,\ldots,x_T)$, the model produces a hidden state $\mathbf{h}_t^{(\ell)}\in\mathbb{R}^{d}$ for token $x_t$ at layer $\ell$. Rather than introducing an additional text encoder, \guard directly takes the concatenated hidden states from a selected subset of layers $\mathcal{S}\subseteq\{1,\ldots,L\}$ as its input. Abstracting the fusion of the selected hidden states into a lightweight prediction head $g_{\theta}$, the token-level output is defined as

\begin{equation}
    \mathbf{s}_t = g_{\theta}\!\left(\left\{\mathbf{h}_t^{(\ell)}\right\}_{\ell\in\mathcal{S}}\right) = \left[\mathbf{s}^{\mathrm{query}}_t,\,s^{\mathrm{unsafe}}_t,\,s^{\mathrm{hallu}}_t\right]\in\mathbb{R}^{\mu + 1 + 1},
\end{equation}

where $\mathbf{s}^{\mathrm{query}}_t\in\mathbb{R}^{\mu}$ contains $\mu$ fine-grained query-intent scores, while $s^{\mathrm{unsafe}}_t\in\mathbb{R}$ and $s^{\mathrm{hallu}}_t\in\mathbb{R}$ respectively represent the response-unsafety and hallucination scores. All scores are produced at every token position. Because each selected hidden state depends only on the current prefix $x_{\leq t}$, \guard provides causally updated predictions throughout generation, enabling token-level streaming detection without an additional encoding pass. We treat these outputs as logits and apply task-specific normalization when probabilities or decisions are required.

\partitle{Query Intent Classification} Following SingGuard~\citep{team2026singguard}, we organize query intents into eight categories and use the first eight outputs of \guard as their class scores: \textbf{(A)} Sexual Content Risk, \textbf{(B)} Real-World Crimes \& Public Safety, \textbf{(C)} Unethical Behavior, \textbf{(D)} Cybersecurity \& Information Manipulation, \textbf{(E)} Agent Safety, \textbf{(F)} Politically Sensitive Content, \textbf{(G)} Animal Abuse, and \textbf{(H)} Safe. The first seven categories cover different types of potentially harmful intent, while Safe indicates that the query does not match any active risk category. A higher class score indicates stronger evidence for the corresponding intent category.

\partitle{Response Safety and Hallucination Classification} The other output scores track risks in the generated response. The response-unsafety score measures the safety risk of the response prefix generated up to the current token, while the hallucination score estimates the likelihood that the generated content contains hallucinations. Higher values indicate that the response is more likely to be unsafe or contain hallucinations, respectively. Both scores are updated throughout decoding, providing streaming risk signals that can support early intervention.

\begin{figure}
    \centering
    \includegraphics[width=0.95\linewidth]{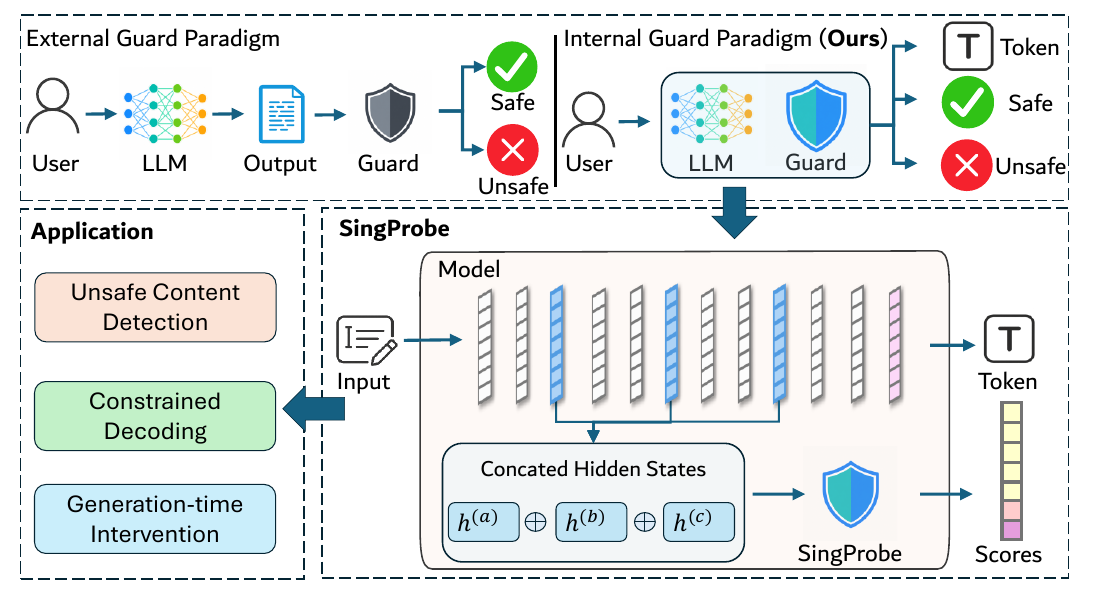}
    \caption{The overview of \guard. \textbf{Top:} SingProbe replaces the conventional external guard paradigm with an intrinsic guard that operates alongside autoregressive decoding. \textbf{Bottom right:} It directly consumes concatenated hidden states from selected base-model layers and produces token-level signals for query intent, response safety, and hallucination risk. \textbf{Bottom left:} These streaming signals support applications including unsafe content detection, constrained decoding, and generation-time intervention.}
    \label{fig:overview}
    \vspace{-1em}
\end{figure}

\subsection{Training Objectives}


We could jointly train \guard for any query- or response-classification task. In our implementation, we train \guard on query-intent classification, response-safety classification, and response-hallucination detection. The three objectives are combined as
\begin{equation}
\mathcal{L}_{\mathrm{total}} = w_q\,\mathcal{L}_{\mathrm{query}} + w_s\,\mathcal{L}_{\mathrm{safety}} + w_h\,\mathcal{L}_{\mathrm{hallu}},
\end{equation}
where $w_q$, $w_s$, and $w_h$ weight the query-intent, response-safety, and response-hallucination objectives, respectively. Rather than keeping these weights fixed, we update them based on exponential moving averages (EMAs) of the corresponding losses, assigning larger weights to tasks with higher recent losses while retaining their configured base weights. Concretely, let $\mathcal{L}_{\tau}^{(t)}$ denote the loss of task $\tau\in\{q,s,h\}$ at step $t$, and $\tilde{\mathcal{L}}_\tau^{(t)}$ its EMA updated with smoothing coefficient $\alpha$:
\begin{equation}
\tilde{\mathcal{L}}_\tau^{(t)} = \alpha\,\tilde{\mathcal{L}}_\tau^{(t-1)} + (1-\alpha)\,\mathcal{L}_\tau^{(t)}, \qquad \tau\in\{q,s,h\}.
\end{equation}

The per-step weight is then obtained by scaling the base weight $w_\tau^{\mathrm{base}}$ with the ratio of the task's EMA loss to the mean of those EMAs, $\bar{\mathcal{L}}^{(t)}=\text{Mean}(\tilde{\mathcal{L}}_\tau^{(t)})$:
\begin{equation}
w_\tau^{(t)} = w_\tau^{\mathrm{base}}\,\frac{\tilde{\mathcal{L}}_\tau^{(t)}}{\bar{\mathcal{L}}^{(t)}}\ , \qquad \tau\in\{q,s,h\}.
\end{equation}

Dividing by $\bar{\mathcal{L}}^{(t)}$ reweights each task in proportion to its recent difficulty (larger recent loss $\Rightarrow$ larger weight), while the three ratios average to~1, so the overall loss magnitude is preserved. We set $\alpha{=}0.9$ and guard the denominator against zero with a small floor of $10^{-9}$. This allows training to focus on the currently more difficult objectives as optimization progresses. The details of the above objectives are introduced as follows.

\subsubsection{Query-Intent Objective}

The query-intent objective applies binary cross-entropy (BCE) independently to the seven risk categories, allowing multiple risks to coexist, and to the Safe category, which is mutually exclusive with all risk categories. In addition to this label-level constraint, we introduce a soft mutual-exclusion loss that penalizes simultaneously high predictions for Safe and the most probable risk category. This encourages a clear separation between safe and risky queries without restricting co-occurrence among different risk types. The query-intent loss is defined as

\begin{equation}
\begin{aligned}
\mathcal{L}_{\mathrm{query}}
={}& \mathbb{E}_{b}\!\Biggl[\frac{1}{|\mathcal{V}_b|}\sum_{t\in\mathcal{V}_b}
\Biggl(
\frac{1}{\mu-1}\sum_{k=0}^{\mu-2}\mathrm{BCE}(\ell_{b,t,k},y_{b,t,k})
+ \lambda_{\mathrm{safe}}\mathrm{BCE}(\ell_{b,t,\mu-1},y_{b,t,\mu-1})
\\
&\qquad
+ \lambda_{\mathrm{mutex}}\,\sigma(\ell_{b,t,\mu-1})
\max_{0\leq k\leq\mu-2}\sigma(\ell_{b,t,k})
\Biggr)\Biggr].
\end{aligned}
\end{equation}

Here, $\mathcal{V}_b$ denotes the valid supervised positions in sample $b$; $\ell_{b,t,k}$ and $y_{b,t,k}$ are the logit and label for category $k$ at position $t$; $\sigma(\cdot)$ is the sigmoid function; and $\lambda_{\mathrm{safe}}$ and $\lambda_{\mathrm{mutex}}$ control the Safe loss and mutual-exclusion penalty, respectively.

\subsubsection{Response Safety and Hallucination Objectives}

Streaming response detection requires learning from every generation prefix, yet available safety and hallucination datasets typically provide only response-level labels rather than fine-grained token-level supervision. Qwen3Guard-Stream~\citep{zhao2025qwen3guard} addresses this problem by generating multiple rollouts from each token prefix and estimating its risk from the proportion of unsafe continuations, followed by LLM-based verification. Although this procedure produces fine-grained boundary labels, repeatedly generating and judging multiple continuations for every prefix incurs substantial annotation cost. Constitutional Classifier++~\citep{cunningham2026CC++} instead avoids explicit token-level annotation through a softmax-weighted objective that concentrates supervision on tokens with high predicted harmfulness. However, because the weights are determined directly by the current logits, this strategy can be sensitive to poorly calibrated predictions at initialization: spuriously high-scoring tokens may receive disproportionate weights and amplify early training noise. 

Based on the above understanding, we therefore adopt an adaptive confidence-weighted aggregator that activates confidence weighting only when token predictions are sufficiently differentiated and otherwise falls back to uniform averaging. The response-safety and response-hallucination objectives are both trained with token-level BCE under this aggregator. Let $r\in\mathcal{R}=\{\mathrm{safety},\mathrm{hallu}\}$ denote the task and $c_r$ its output dimension, where $c_{\mathrm{safety}}=\mu$ and $c_{\mathrm{hallu}}=\mu+1$. The corresponding objective is defined as

\begin{equation}
\begin{aligned}
\mathcal{L}_{r}
&= \mathbb{E}_{b}\!\left[\sum_{t\in\mathcal{V}_b}\alpha^{(r)}_{b,t}\,
\mathrm{BCE}(\ell_{b,t,c_r},y_{b,t,c_r})\right],
\\
\alpha^{(r)}_{b,t}
&=\operatorname{softmax}_{t\in\mathcal{V}_b}\!\left(\frac{\beta_r\,p^{(r)}_{b,t}}{T}\right),
\qquad
\beta_r=\operatorname{clip}\!\left(\frac{\operatorname{std}_{(b,t)\in\mathcal{V}}(p^{(r)}_{b,t})}{\tau},0,1\right),
\quad p^{(r)}_{b,t}=\sigma(\ell_{b,t,c_r}).
\end{aligned}
\end{equation}

Here, $p^{(r)}_{b,t}$ is the predicted risk probability and $\alpha^{(r)}_{b,t}$ is its normalized token weight. The factor $\beta_r$ controls the strength of confidence weighting according to the dispersion of valid predictions. When predictions are insufficiently differentiated, $\beta_r$ approaches zero and the aggregator reduces to uniform averaging, preventing early training noise from being amplified. Once the predictions become sufficiently differentiated, $\beta_r$ increases and the softmax assigns greater weight to tokens with higher predicted risk---unsafety for response safety and hallucination probability for hallucination detection. $T$ is the weighting temperature, and $\tau$ controls the sensitivity of the adaptive factor.

This confidence weighting has complementary effects on negative and positive samples. For safe or non-hallucinated responses, high-risk tokens correspond to potential false positives; emphasizing their losses directly suppresses spurious risk peaks. For unsafe or hallucinated responses, tokens in a benign prefix typically receive lower risk scores and are therefore downweighted relative to the later risk-bearing tokens. This prevents the response-level positive label from incorrectly pushing every prefix toward the risky class. Moreover, because the weights are normalized separately within each sample, a benign prefix receives a larger relative weight in a negative response, where no genuinely risky continuation dominates the weighting distribution, than in a positive response containing later high-risk tokens. Consequently, similar benign prefixes receive stronger supervision toward the safe or non-hallucinated class from negative samples while remaining largely unaffected by the coarse positive labels of unsafe or hallucinated samples.

All objectives are computed only at response positions with valid labels. Because token weights are normalized within each sample, longer responses do not dominate training.

\subsection{Training Dataset Construction}
\label{sec:training_corpus}

Training \guard requires stable supervision for learning safety- and factuality-related decision boundaries in the hidden-state space of the backbone LLM. We therefore construct a unified corpus for three tasks: \emph{query intent classification}, \emph{response safety classification}, and \emph{response hallucination detection}. 
The final corpus covers harmful prompts, benign but sensitive prompts, harmful responses, safe refusals, and helpful safe completions. Each example follows a unified format containing a user query, optional dialogue context, and an optional assistant response. Query-side and response-side labels are maintained separately, allowing the same data source to support prompt screening, response monitoring, and joint query--response classification. The corpus is constructed through open-source data curation, policy-grounded synthesis, and multi-stage quality control.

\partitle{Open-Source Data Curation}
We use only the official training splits of publicly available safety benchmarks and training datasets. Because different sources adopt different annotation protocols, risk granularities, and decision thresholds, directly merging their labels would introduce inconsistent supervision. We therefore normalize all safety examples under the fine-grained risk taxonomy introduced in this work.

For datasets with existing annotations, the original labels are first mapped to our taxonomy using predefined category correspondences and then independently verified by an LLM judge under our safety definitions. Samples with consistent labels are retained, whereas ambiguous cases, including potential safe/unsafe polarity flips, are moved to an unlabeled hard-case pool. For unlabeled data and hard cases, multiple open-source LLMs, including Qwen3.5-397B-A17B~\citep{qwen3.5} and Kimi-K2.6~\citep{kimik26}, independently annotate each example. We combine self-consistency across repeated predictions from the same model with agreement across different models. Each accepted sample must pass both a binary safe/unsafe consistency check and a fine-grained category check. Samples failing the binary check are discarded, while samples with unresolved category disagreement are retained only after targeted review. 


For hallucination supervision, we additionally incorporate the training splits of public factuality and hallucination datasets, labeling correct answers and honest refusals as \emph{non-hallucinated} and incorrect, fabricated, or unsupported answers as \emph{hallucinated}.

\partitle{Policy-Grounded Safety Data Synthesis}
Open-source safety data are typically long-tailed and provide limited coverage of emerging or low-frequency risks. We therefore supplement them with policy-grounded synthetic data. For each risk category, an internally red-teamed jailbreak model generates harmful prompts, while an aligned model produces topically similar but benign counterparts. These contrastive pairs share surface vocabulary but differ in their underlying intent, discouraging \guard from relying on lexical shortcuts for the safety task.

We further generate multiple responses for each prompt. Unsafe prompts are paired with both harmful responses and policy-compliant refusals or safe alternatives, while benign prompts are paired with helpful and compliant responses. This construction prevents response safety from being inferred directly from the prompt label and encourages \guard to track the evolving response representation. We additionally synthesize multi-category and multilingual examples to improve coverage of compositional and cross-lingual risks. All synthetic samples are independently re-evaluated by an LLM that is not involved in generation, and samples whose predicted labels disagree with the intended labels are removed.

\partitle{Quality Control and Training Supervision}
We remove duplicate and near-duplicate examples, balance safe and unsafe instances across major risk categories, and retain high-confidence boundary cases that distinguish harmful intent from benign discussions of sensitive topics. During training, the curated examples are processed by the backbone LLM to obtain the hidden states used by \guard. Query labels are aligned with representations at the query boundary, while response-safety and hallucination labels supervise representations produced along the response trajectory. This formulation enables \guard to jointly monitor user intent, response safety, and factual reliability without requiring an additional text encoder or a separately deployed guardrail model.

\section{SingStreamBench: Benchmarking Streaming Safety Detection}

\subsection{Motivation}

Existing safety benchmarks are primarily designed for post-hoc classification~\citep{ji2023beavertails, han2024wildguard,
mazeika2024harmbench, ghosh2025aegis2} and provide only a single response-level label. Such labels indicate whether a completed response is unsafe, but do not reveal when unsafe content first emerges during generation. Consequently, they cannot adequately evaluate whether a streaming guardrail remains silent on benign prefixes, detects the actual onset of harmful content, or intervenes with minimal delay.

FineHarm~\citep{li2025fineharm} takes an important step toward fine-grained evaluation by providing token-level harmfulness annotations. Its annotations are constructed heuristically: responses are first labeled at the sentence level using an LLM, after which content words in harmful sentences are marked as harmful and their labels are mapped to the corresponding tokens. However, our manual inspection reveals substantial noise in these fine-grained labels. More fundamentally, harmfulness is often expressed compositionally and depends on surrounding context, making it difficult to assign a stable safety label to an individual token in isolation. These issues limit the reliability of token-level labels as ground truth for evaluating the precise onset of unsafe content.

Even when fine-grained labels can be obtained, the composition of existing safety datasets introduces another limitation. Most unsafe responses begin producing harmful content near the start of the response, resulting in a strong positional bias and few examples containing a substantial safe prefix before the unsafe segment. A classifier may therefore appear effective by reacting to the unsafe query or early lexical cues, without demonstrating that it can remain inactive throughout benign content and trigger only when harmful content actually appears.

Based on these observations, we construct a manually verified fine-grained benchmark, \textbf{SingStreamBench}, and a larger-scale version, \textbf{SingStreamBench-Full}, to evaluate both the accuracy and timeliness of streaming safety detection. Instead of assigning intrinsic labels to isolated tokens, we segment each response into sentences and annotate its cumulative sentence-level prefixes to identify a semantically meaningful unsafe onset boundary. We further construct controlled safe-to-unsafe transitions and fully safe counterexamples, organized into six tiers that vary the complexity and safety composition of both queries and response prefixes. This design enables explicit evaluation of whether a guardrail remains silent before risk emerges, detects the true unsafe segment promptly, and avoids relying on spurious query or positional cues.

\subsection{Benchmark Design}

Each benchmark sample consists of a query $\mathbf{q}$ and a response $\mathbf{r}$ that can be conceptually decomposed as
\begin{equation}
\mathbf{r}=\mathbf{r}^{\mathrm{safe}}\oplus\mathbf{r}^{\mathrm{target}},
\end{equation}
where $\mathbf{r}^{\mathrm{safe}}$ is an optional safe prefix and $\mathbf{r}^{\mathrm{target}}$ is the continuation whose safety status determines the response label. The central evaluation target is the transition around the beginning of $\mathbf{r}^{\mathrm{target}}$: a reliable streaming guardrail should remain silent throughout the safe prefix and react promptly once unsafe content emerges. For fully safe responses, it should remain silent throughout generation.

We organize the benchmark into six tiers, as summarized in Table~\ref{tab:singstreambench-tiers}. The tiers progressively test spurious associations with the query or safe prefix and robustness to increasingly complex contexts.

\begin{table}[t]
\centering
\small
\caption{Tiered construction of SingStreamBench.}
\label{tab:singstreambench-tiers}
\begin{tabular}{@{}c p{0.20\linewidth} p{0.27\linewidth} p{0.43\linewidth}@{}}
\toprule
\textbf{Tier} & \textbf{Query} & \textbf{Response structure} & \textbf{Primary purpose} \\
\midrule
0 & Original query & Original response & Establish performance on ordinary query--response samples. \\
1 & Unsafe query & Safe refusal & Test whether the guardrail spuriously reacts to unsafe intent in the query despite a safe response. \\
2 & Original query & Safe prefix $+$ safe continuation & Test whether the guardrail remains silent when a safe prefix is continued safely. \\
3 & Original query & Complex safe prefix $+$ unsafe response & Test whether background information or disclaimers delay or otherwise interfere with detection at the true unsafe onset. \\
4 & Background context $+$ original query & Original unsafe response & Test robustness to complex query-side context and whether the guardrail attributes response risk correctly. \\
5 & Multi-question query & Safe answers/prefix $+$ unsafe response & Jointly test complex query context and a substantial safe response prefix before unsafe content appears. \\
\bottomrule
\end{tabular}
\end{table}

Tiers 1 and 2 primarily diagnose false associations. Tier 1 tests whether an unsafe query alone causes the response stream to be flagged even when the model safely refuses, while Tier 2 tests whether a safe prefix itself induces false alarms when followed by a safe continuation. Tiers 3--5 introduce more complex contexts. Tier 3 inserts background knowledge or a disclaimer before the original unsafe response; Tier 4 prepends query-related background text to the query; and Tier 5 combines one unsafe question with multiple safe questions, producing a multi-question context in which safe content precedes the unsafe segment. Together, these tiers distinguish genuine localization of response risk from shortcuts based on query harmfulness, response position, or superficial lexical cues.

\subsection{Construction Pipeline}

The benchmark is constructed through the following four-stage pipeline.

\begin{enumerate}[leftmargin=*]
    \item \textbf{Seed sampling and verification.} We sample query--response pairs from existing safety datasets and re-evaluate whether each response-level label is correct. Three judge models, Qwen3-235B-A22B~\citep{yang2025qwen3}, GLM-5.2~\citep{zeng2026glm5}, and Kimi-K2.6~\citep{kimik26}, independently verify each sample, and only samples unanimously accepted by all three models are retained. This step removes mislabeled and ambiguous seed responses.

    \item \textbf{Tiered data augmentation.} Starting from a verified unsafe query--response pair, we construct four types of augmented samples. For Tier 1, a safe refusal is generated for the unsafe query. For Tier 3, the model generates safe background information or a disclaimer that is prepended to the original unsafe response. For Tier 4, query-related background text is generated and prepended to the original query. For Tier 5, one unsafe question is combined with one to three sampled safe questions using a multi-question template such as ``Please answer the following questions: 1.~\ldots; 2.~\ldots''. For each generation operation, one of the three models above is selected at random to increase stylistic and model diversity.

    \item \textbf{Sentence-boundary prefix annotation.} Each response is first segmented into sentences $\mathbf{u}_1,\ldots,\mathbf{u}_m$. We then construct cumulative prefixes $\mathbf{P}_j=\mathbf{u}_1\oplus\cdots\oplus\mathbf{u}_j$ and use Kimi-K2.6 to assign a Safe or Unsafe label to each prefix $\mathbf{P}_j$, rather than labeling each sentence in isolation. For an unsafe response, annotation proceeds sequentially and stops once the first unsafe prefix is identified; the starting position of its newly added sentence is recorded as \texttt{Unsafe\_Start\_Index}. All preceding prefixes are treated as safe. For a safe response, every sentence-level prefix is verified as safe. This procedure directly identifies the safe-prefix boundary required for streaming evaluation while avoiding token-level semantic ambiguity.

    \item \textbf{Safe-continuation augmentation.} To construct Tier 2, we sample examples with a non-empty safe prefix and ask a model to continue generation safely from the corresponding query and prefix. The resulting fully safe responses test whether a guardrail can remain inactive after observing prefixes that may otherwise also occur before unsafe continuations.

\end{enumerate}

\subsection{Dataset Variants and Statistics}

The construction pipeline produces two benchmark variants. \textbf{SingStreamBench} is the high-quality core benchmark containing 210 samples that have undergone manual verification. \textbf{SingStreamBench-Full} is the larger-scale variant containing 2,428 constructed samples, providing broader coverage across tiers and safety scenarios. The original queries and responses are partly from BeaverTails~\citep{ji2023beavertails}, PKU-SafeRLHF~\citep{ji2024pku}, WildGuard~\citep{han2024wildguard}, XSTest~\citep{rottger2023xstest}, and expguard~\citep{choi2026expguard}. The manually verified core set supports reliable comparison of streaming detection accuracy and latency, while the full set enables evaluation at a larger scale.

\section{Evaluation}

\subsection{Experimental Settings}

\partitle{Safety Guard Baselines} We compare \guard against two proprietary LLMs and 12 open-source guardrail configurations. The proprietary baselines are Gemini 3 Pro~\citep{google2025gemini3} and GPT-5.1~\citep{openai2025gpt5}. The open-source baselines include YuFeng-XGuard-Reason-0.6B/8B~\citep{lin2026yufengxguard}, Llama Guard 3~\citep{inan2023llamaguard}, WildGuard~\citep{han2024wildguard}, GraniteGuardian~\citep{graniteguardian2024}, ShieldGemma-9B~\citep{zeng2024shieldgemma}, and Qwen3Guard series~\citep{zhao2025qwen3guard}. For Qwen3Guard, we evaluate both loose and strict modes, which respectively treat the Controversial label as safe and unsafe. Some of the baselines' evaluation results are from \citet{team2026singguard}. We additionally implement three intrinsic-guard baselines: \emph{Vanilla}, which uses standard uniformly averaged BCE; \emph{CC++}, which reproduces Anthropic's Constitutional Classifier++ objective~\citep{cunningham2026CC++}; and \emph{Kelp}~\citep{li2025kelp}. To ensure a fair comparison, all three baselines use the same training datasets, Ling-3.0-flash base model, and model architecture as \guard. Because these methods are designed specifically for response safety, we compare them only on response-safety evaluations.

\partitle{Hallucination Detection Baselines} We compare \guard with four representative hidden-state-based hallucination detectors: DRIFT~\citep{bhatnagar2026drift}, HaMI~\citep{niu2025robust}, and SAPLMA~\citep{azaria2023internal}. These methods cover lightweight probing, adaptive token selection, and training-free NTK-based scoring. For SAPLMA, we evaluate both the raw-input and response-formatted variants, denoted SAPLMA-raw and SAPLMA-response, respectively.

\partitle{Implementation Details of \guard} In our experiments, we implement a unified \guard architecture based on a query-residual multi-head attention (MHA) block. Specifically, the block computes queries $Q$, keys $K$, and values $V$ and produces the attention output $O$, and finally outputs $\mathrm{Norm}(Q + O)$, followed by a linear classifier that produces the guardrail scores. As the default setting, we use hidden states from three layers of the base model as input to \guard, with the layers evenly selected from the shallow, middle, and deep portions of the model. Unless otherwise stated, the main-text experiments use Ling-3.0-tiny-singprobe and Ling-3.0-flash-singprobe. We additionally extend \guard to 29 open-source base models spanning diverse model families, parameter scales, and architectures; the additional model-wise results are provided in Appendix~\ref{app:cross_model_evaluation}.

\subsection{Query Classification Evaluation}

\begin{table}[t]
    \centering
    \caption{Query intent classification evaluation results. The values in the table are F1 scores.}
    \vspace{0.5mm}
    \label{tab:text_query_detail}
    \resizebox{\textwidth}{!}{%
    \renewcommand{\arraystretch}{1.10}
    \setlength{\tabcolsep}{3.8pt}
    \begin{tabular}{l c c c c c c >{\columncolor{green!8}}c}
        \toprule[1.2pt]
        \textbf{Model}
        & \textbf{Aegis2}
        & \textbf{XGuard Test}
        & \textbf{OpenAI Moderation}
        & \textbf{XSTest}
        & \textbf{HarmBench}
        & \textbf{ExpGuardTest}
        & \cellcolor{green!8}\textbf{Avg.} \\
        \midrule[0.8pt]
        Gemini3-Pro
        & 0.7401 & 0.8426 & 0.7545 & 0.8962 & 0.8615 & 0.7980
        & \cellcolor{green!8}0.8155 \\
        
        GPT-5.1
        & 0.8142 & 0.8815 & 0.8024 & 0.9169 & 0.9392 & 0.8554
        & \cellcolor{green!8}0.8683 \\

        YuFeng-XGuard-Reason-0.6B
        & 0.8625 & 0.9126 & 0.7107 & 0.9179 & 0.9250 & 0.7476
        & \cellcolor{green!8}0.8461 \\
        
        YuFeng-XGuard-Reason-8B
        & 0.8541 & \textbf{0.9266} & 0.7187 & 0.9461 & \textbf{0.9466} & 0.8361
        & \cellcolor{green!8}\textbf{0.8714} \\
        
        Llama Guard 3
        & 0.7635 & 0.7909 & 0.7904 & 0.8852 & 0.8544 & 0.7077
        & \cellcolor{green!8}0.7987 \\
        
        WildGuard
        & 0.8076 & 0.8237 & 0.7252 & \textbf{0.9501} & 0.9288 & 0.8438
        & \cellcolor{green!8}0.8465 \\
        
        GraniteGuardian
        & 0.7138 & 0.5777 & 0.5684 & 0.7269 & 0.9045 & \textbf{0.9115}
        & \cellcolor{green!8}0.7338 \\
        
        ShieldGemma-9B
        & 0.7588 & 0.6885 & \textbf{0.8143} & 0.8293 & 0.6441 & 0.5340
        & \cellcolor{green!8}0.7115 \\

        Qwen3Guard-Gen-4B-loose
        & 0.8054 & 0.7952 & 0.7897 & 0.8674 & 0.8311 & 0.7139
        & \cellcolor{green!8}0.8005 \\

        Qwen3Guard-Gen-4B-strict
        & 0.8499 & 0.8789 & 0.6764 & 0.8791 & 0.9260 & 0.8312
        & \cellcolor{green!8}0.8402 \\
        
        Qwen3Guard-Gen-8B-loose
        & 0.8199 & 0.8197 & 0.7958 & 0.8815 & 0.8401 & 0.6953
        & \cellcolor{green!8}0.8087 \\
        
        Qwen3Guard-Gen-8B-strict
        & 0.8498 & 0.8797 & 0.6804 & 0.8796 & 0.9326 & 0.8297
        & \cellcolor{green!8}0.8420 \\

        Qwen3Guard-Stream-8B-loose
        & 0.8091 & 0.8282 & 0.8023 & 0.9016 & 0.8390 & 0.7519
        & \cellcolor{green!8}0.8220 \\
        
        Qwen3Guard-Stream-8B-strict
        & \textbf{0.8602} & 0.8752 & 0.7386 & 0.9223 & 0.9231 & 0.8416
        & \cellcolor{green!8}0.8602 \\
        
        \midrule[0.8pt]
        \textbf{Ling-3.0-tiny-singprobe} & 0.8488 & 0.8684 & 0.7510 & 0.8926 & 0.9035 & 0.8724 & 0.8561 \\
        \textbf{Ling-3.0-flash-singprobe} & 0.8481 & 0.8795 & 0.7703 & 0.9012 & 0.9231 & 0.8825 & 0.8674 \\
        \bottomrule[1.2pt]
    \end{tabular}
    }
\end{table}

Table~\ref{tab:text_query_detail} reports the query-intent classification results across six benchmarks, including Aegis2~\citep{ghosh2025aegis2}, XGuard~\citep{lin2026yufengxguard}, OpenAI Moderation~\citep{markov2023openaimoderation}, XSTest~\citep{rottger2023xstest}, HarmBench~\citep{mazeika2024harmbench}, and ExpGuard~\citep{choi2026expguard}. For each sample, we feed the query to the base model and let it generate a response with a maximum length of 512. \guard produces query-intent scores at every token position in the generated response. We then average these scores across response positions and use the resulting eight-dimensional score vector as the final query-intent prediction. The best configuration, \guard with Ling-3.0-flash, achieves an average F1 of 0.8674, ranking closely behind YuFeng-XGuard-Reason-8B (0.8714) and GPT-5.1 (0.8683). It also outperforms all evaluated Qwen3Guard configurations, whose best average F1 is 0.8602, as well as other widely used guardrails such as WildGuard and Llama Guard 3. These results show that an intrinsic lightweight head can provide query-safety classification performance comparable to substantially larger standalone guardrail models and frontier proprietary LLMs.

The results further demonstrate consistent benefits from scaling the base model. Replacing Ling-3.0-tiny with Ling-3.0-flash improves the average F1 of \guard from 0.8561 to 0.8674, indicating that stronger base-model hidden states provide richer query-safety signals for the lightweight head. \guard is competitive with the strongest standalone guardrails across the six benchmarks and achieves the best F1 on ExpGuardTest (0.8825), suggesting that the base model's hidden states already contain strong query-safety signals that can be effectively extracted by a lightweight temporal head.

\subsection{Response Safety Evaluation}

We evaluate response safety from three complementary dimensions: \emph{response-level classification}, which measures whether a guardrail correctly identifies the overall safety of a completed response; \emph{streaming detection}, which evaluates both response-level discrimination and token-level risk localization throughout generation; and \emph{false-positive robustness}, which measures whether the guardrail remains silent on nominally benign queries and freely generated responses.

\partitle{Response-level Classification} We first evaluate response safety under teacher forcing by feeding each query together with its labeled response. We evaluate 8 benchmarks including BeaverTails~\citep{ji2023beavertails}, PKU-SafeRLHF~\citep{ji2024pku}, Aegis2~\citep{ghosh2025aegis2}, WildGuard~\citep{han2024wildguard}, XGuard~\citep{lin2026yufengxguard}, HarmBench~\citep{mazeika2024harmbench}, XSTest~\citep{rottger2023xstest}, and ExpGuard~\citep{choi2026expguard}. \guard produces a response-safety score at every response position, and we use the maximum score over the entire response as the final response-level risk score. As shown in Table~\ref{tab:text_response_detail}, \guard with Ling-3.0-flash achieves the best average F1 of 0.8728, outperforming the strongest standalone baseline, Qwen3Guard-Gen-8B-strict, by 1.24 points. Under the matched intrinsic-guard setting with Ling-3.0-flash, it also exceeds Vanilla (0.8535), CC++ (0.8652), and Kelp (0.8630) by 1.93, 0.76, and 0.98 points, respectively. \guard with Ling-3.0-tiny remains competitive at 0.8508, and scaling from Ling-3.0-tiny to Ling-3.0-flash yields consistent gains across nearly all benchmarks. In particular, \guard with Ling-3.0-flash achieves the best F1 on WildGuard (0.8000) and XGuard Test (0.8588), while CC++ obtains the highest score on ExpGuardTest (0.9253).

\begin{table}[t]
    \centering
    \caption{Response safety classification evaluation results. The values in the table are F1 scores.}
    \vspace{0.5mm}
    \label{tab:text_response_detail}
    \resizebox{\textwidth}{!}{%
    \renewcommand{\arraystretch}{1.10}
    \setlength{\tabcolsep}{4.2pt}
    \begin{tabular}{l c c c c c c c c >{\columncolor{green!8}}c}
        \toprule[1.2pt]
        \textbf{Model} & \textbf{BeaverTails} & \textbf{PKU-SafeRLHF} & \textbf{Aegis2} & \textbf{WildGuard} & \textbf{XGuard Test} & \textbf{HarmBench} & \textbf{XSTest} & \textbf{ExpGuardTest} & \cellcolor{green!8}\textbf{Avg.} \\
        \midrule[0.8pt]
        \textbf{Externel Guard} \\
        Gemini3-Pro & 0.7960 & 0.8605 & 0.7667 & 0.5478 & 0.7172 & 0.7632 & 0.5984 & 0.8921 & \cellcolor{green!8}0.7427 \\
        GPT-5.1 & 0.8042 & 0.8880 & 0.7861 & 0.6723 & 0.7292 & 0.7812 & 0.6667 & 0.9151 & \cellcolor{green!8}0.7804 \\
        
        YuFeng-XGuard-Reason-0.6B    & 0.8624 & 0.9181 & 0.8253 & 0.7471 & 0.8547 & 0.8759 & 0.8000 & 0.9125 & \cellcolor{green!8}0.8495 \\
        YuFeng-XGuard-Reason-8B & 0.8571 & 0.9144 & 0.8303 & 0.7350 & 0.8556 & 0.8672 & 0.8387 & 0.9121 & \cellcolor{green!8}0.8513 \\
        Llama Guard 3 & 0.6778 & 0.8908 & 0.6110 & 0.7030 & 0.6667 & 0.8668 & 0.9041 & 0.8419 & \cellcolor{green!8}0.7703 \\
        WildGuard & 0.7789 & 0.8189 & \textbf{0.8644} & 0.5194 & 0.6265 & 0.6976 & 0.5929 & 0.8998 & \cellcolor{green!8}0.7248 \\
        GraniteGuardian & 0.7291 & 0.6869 & 0.6324 & 0.2861 & 0.6837 & 0.6667 & 0.2977 & 0.8868 & \cellcolor{green!8}0.6087 \\
        ShieldGemma-9B & 0.6962 & 0.8214 & 0.6765 & 0.5316 & 0.6142 & 0.6463 & 0.8652 & 0.4014 & \cellcolor{green!8}0.6566 \\
        Qwen3Guard-Gen-4B-loose   & 0.8455 & \textbf{0.9317} & 0.7874 & 0.7850 & 0.7254 & 0.8870 & \textbf{0.9172} & 0.8230 & \cellcolor{green!8}0.8378 \\
        Qwen3Guard-Gen-4B-strict  & 0.8581 & 0.9060 & 0.8351 & 0.7750 & 0.7953 & \textbf{0.9061} & 0.8721 & 0.8986 & \cellcolor{green!8}0.8558 \\
        Qwen3Guard-Gen-8B-loose & 0.8493 & 0.9308 & 0.8309 & 0.7881 & 0.7542 & 0.8816 & 0.9103 & 0.8455 & \cellcolor{green!8}0.8488 \\
        Qwen3Guard-Gen-8B-strict & \textbf{0.8669} & 0.9043 & 0.8407 & 0.7778 & 0.7906 & 0.8907 & 0.8982 & 0.9140 & \cellcolor{green!8}0.8604 \\
        Qwen3Guard-Stream-8B-loose & 0.8560 & 0.9076 & 0.8203 & 0.7866 & 0.7651 & 0.8746 &  0.8539 & 0.8973 & \cellcolor{green!8}0.8452 \\
        Qwen3Guard-Stream-8B-strict & 0.8570 & 0.9019 & 0.8208 & 0.7713 & 0.7697 & 0.8882 & 0.8556 & 0.9244 & \cellcolor{green!8}0.8486 \\
        \midrule[0.8pt]
        \textbf{Intrinsic Guard} \\
        Ling-3.0-flash-Vanilla & 0.8377 & 0.9165 & 0.8247 & 0.7854 & 0.8215 & 0.8690 & 0.8521 & 0.9212 & 0.8535 \\
        Ling-3.0-flash-CC++ & 0.8499 & 0.9218 & 0.8389 & 0.7769 & 0.8382 & 0.8865 & 0.8837 & \textbf{0.9253} & 0.8652 \\
        Ling-3.0-flash-Kelp & 0.8462 & 0.9291 & 0.8219 & 0.7933 & 0.8148 & 0.8746 & 0.9024 & 0.9217 & 0.8630 \\
        \midrule[0.8pt]
        \textbf{Ling-3.0-tiny-singprobe} & 0.8537 & 0.9120 & 0.8211 & 0.7683 & 0.8122 & 0.8645 & 0.8588 & 0.9159 & 0.8508\\
        \textbf{Ling-3.0-flash-singprobe} & 0.8433 & 0.9301 & 0.8333 & \textbf{0.8000} & \textbf{0.8588} & 0.8995 & 0.8929 & 0.9247 & \textbf{0.8728}\\
        \bottomrule[1.2pt]
    \end{tabular}
    }
\end{table}

\partitle{Streaming Safety Detection} Table~\ref{tab:streaming_safety_detail} evaluates detection throughout generation on SingStreamBench, SingStreamBench-Full, and FineHarm~\citep{li2025fineharm}. Response-level AUC (R-AUC) is computed using the maximum response-safety score over all token positions and takes the response-level label as the ground-truth label, whereas token-level AUC (T-AUC) directly evaluates the score at each token position against the corresponding fine-grained annotation. The tokens in the annotated safe prefix are labeled as $0$. Since streaming detection primarily concerns token-level risk localization, we use average T-AUC as the principal metric. All intrinsic-guard configurations outperform the Qwen3Guard-Stream baselines in both average R-AUC and T-AUC. The best streaming configuration is Ling-3.0-flash-singprobe, which obtains the highest average T-AUC of 0.9481 together with an R-AUC of 0.9887. Ling-3.0-tiny-singprobe obtains a comparable 0.9888/0.9479 R-AUC/T-AUC. Relative to the strongest Qwen3Guard-Stream baseline, Ling-3.0-flash-singprobe improves average T-AUC by 4.93 percentage points, while Kelp achieves the highest average R-AUC of 0.9912, a 2.72-point improvement. In the matched intrinsic-guard comparison, Vanilla obtains 0.9739/0.9228 R-AUC/T-AUC, CC++ obtains 0.9875/0.9478, and Kelp obtains 0.9912/0.9193. Thus, the SingProbe objectives substantially improve both response- and token-level detection over uniform BCE. Compared with CC++, Ling-3.0-flash-singprobe further improves average R-AUC and T-AUC to 0.9887 and 0.9481, respectively; Kelp instead prioritizes response-level discrimination at the cost of substantially lower token-level localization. The improvements over external streaming baselines are especially pronounced on SingStreamBench and SingStreamBench-Full, while performance remains comparable on FineHarm.

\begin{table}[t]
    \centering
    \caption{Streaming safety evaluation results. Response-level AUC (R-AUC) evaluates response-wise discrimination using the maximum response-safety score over all token positions as the response risk score. Token-level AUC (T-AUC) evaluates token-wise discrimination by comparing each token's response-safety score with its corresponding fine-grained label.}
    \vspace{0.5mm}
    \label{tab:streaming_safety_detail}
    \resizebox{0.95\textwidth}{!}{%
    \renewcommand{\arraystretch}{1.10}
    \setlength{\tabcolsep}{4.2pt}
    \begin{tabular}{l c c c c c c >{\columncolor{green!8}}c >{\columncolor{green!8}}c}
        \toprule[1.2pt]
        \multirow{2}{*}{\textbf{Model}}
        & \multicolumn{2}{c}{\textbf{SingStreamBench}}
        & \multicolumn{2}{c}{\textbf{SingStreamBench-Full}}
        & \multicolumn{2}{c}{\textbf{FineHarm}}
        & \multicolumn{2}{>{\columncolor{green!8}}c}{\textbf{Avg.}} \\
        \cmidrule(lr){2-3}
        \cmidrule(lr){4-5}
        \cmidrule(lr){6-7}
        \cmidrule(lr){8-9}
        & \textbf{R-AUC}
        & \textbf{T-AUC}
        & \textbf{R-AUC}
        & \textbf{T-AUC}
        & \textbf{R-AUC}
        & \textbf{T-AUC}
        & \textbf{R-AUC}
        & \textbf{T-AUC} \\
        \midrule[0.8pt]
        \textbf{Externel Guard}\\

        Qwen3Guard-Stream-4B-loose
        & 0.9594 & 0.8849
        & 0.9194 & 0.8389
        & 0.9874 & 0.9456
        & 0.9554 & 0.8898 \\

        Qwen3Guard-Stream-4B-strict
        & 0.9669 & 0.8896
        & 0.9313 & 0.8528
        & 0.9938 & 0.9540
        & 0.9640 & 0.8988 \\

        Qwen3Guard-Stream-8B-loose
        & 0.9591 & 0.8678
        & 0.9239 & 0.8283
        & 0.9905 & 0.9504
        & 0.9578 & 0.8822 \\

        Qwen3Guard-Stream-8B-strict
        & 0.9616 & 0.8684
        & 0.9370 & 0.8447
        & 0.9933 & 0.9549
        & 0.9640 & 0.8893 \\
        \midrule[0.8pt]
        \textbf{Intrinsic Guard} \\
        Ling-3.0-flash-Vanilla & 0.9676 & 0.9175 & 0.9600 & 0.9043 & 0.9940 & 0.9467 & 0.9739 & 0.9228 \\
        Ling-3.0-flash-CC++ & 0.9880 & 0.9533 & 0.9790 & \textbf{0.9344} & 0.9956 & 0.9557 & 0.9875 & 0.9478\\
        Ling-3.0-flash-Kelp & \textbf{0.9936} & 0.9090 & \textbf{0.9844} & 0.8981 & 0.9955 & 0.9509 & \textbf{0.9912} & 0.9193\\
        \midrule[0.8pt]
        \textbf{Ling-3.0-tiny-singprobe} & 0.9935 & 0.9529
          & 0.9775 & 0.9340
          & 0.9954 & \textbf{0.9568}
          & 0.9888 & 0.9479 \\
        \textbf{Ling-3.0-flash-singprobe} & 0.9922 & \textbf{0.9599} & 0.9778 & 0.9294 & \textbf{0.9961} & 0.9549 & 0.9887 & \textbf{0.9481} \\




        \bottomrule[1.2pt]
    \end{tabular}
    }
\end{table}

\begin{table}[t]
  \centering
  \caption{The false positive rate (FPR, in \%) of guardrail models on benign data. Each cell shows \texttt{query/response} FPR (lower is better), and Avg. is the macro average over the five datasets. For Qwen3Guard models we report both the \textit{Loose} (Unsafe
  only) and \textit{Strict} (Unsafe+Controversial) judgments.}
  \label{tab:fpr}
  \small
  \resizebox{0.95\textwidth}{!}{%
    \renewcommand{\arraystretch}{1.10}
    \setlength{\tabcolsep}{4.2pt}
  \begin{tabular}{l ccccc >{\columncolor{green!8}}c}
  \toprule
  Model
   & MATH & databricks-dolly-15k & GSM8K & MBPP & openbookqa & Macro Avg. \\
  \midrule
  \textbf{Externel Guard} \\
  Llama-Guard-3-8B
    & 0.00/0.00
    & 0.25/0.16
    & 0.00/0.00
    & 0.00/0.00
    & 0.05/0.20
    & 0.06/0.07 \\

  YuFeng-XGuard-Reason-8B
    & 0.01/0.01
    & 0.38/0.50
    & 0.15/0.08
    & 0.00/0.00
    & 0.62/0.55
    & 0.23/0.23 \\

  Granite-Guardian-4.1-8B
    & 0.00/0.00
    & 0.37/0.34
    & 0.00/0.00
    & 0.00/0.42
    & 2.86/0.97
    & 0.65/0.35 \\

  ShieldGemma-9B
    & 0.00/0.00
    & 0.27/0.22
    & 0.00/0.00
    & 0.00/0.00
    & 0.89/0.50
    & 0.23/0.14 \\

  WildGuard
    & 0.00/0.00
    & 0.57/0.40
    & 0.00/0.00
    & 0.00/0.00
    & 1.02/0.59
    & 0.32/0.20 \\

    Qwen3Guard-Gen-4B-Loose
    & 0.00/0.00
    & 0.13/0.25
    & 0.00/0.00
    & 0.00/0.00
    & 0.02/0.08
    & 0.03/0.07 \\

  Qwen3Guard-Gen-4B-Strict
    & 0.00/0.00
    & 0.63/0.76
    & 0.00/0.00
    & 0.00/0.00
    & 0.29/0.40
    & 0.18/0.23 \\

  Qwen3Guard-Gen-8B-Loose
    & 0.00/0.00
    & 0.09/0.20
    & 0.00/0.00
    & 0.00/0.00
    & 0.00/0.10
    & 0.02/0.06 \\

  Qwen3Guard-Gen-8B-Strict
    & 0.00/0.00
    & 0.50/0.74
    & 0.00/0.00
    & 0.00/0.00
    & 0.29/0.59
    & 0.16/0.27 \\

  Qwen3Guard-Stream-4B-Loose
    & 0.00/0.00
    & 0.15/0.24
    & 0.00/0.00
    & 0.00/0.00
    & 0.05/0.05
    & 0.04/0.06 \\

  Qwen3Guard-Stream-4B-Strict
    & 0.00/0.00
    & 0.76/0.28
    & 0.00/0.00
    & 0.00/0.00
    & 0.32/0.05
    & 0.22/0.07 \\

  Qwen3Guard-Stream-8B-Loose
    & 0.00/0.00
    & 0.12/0.23
    & 0.00/0.00
    & 0.00/0.00
    & 0.00/0.02
    & 0.02/0.05 \\

  Qwen3Guard-Stream-8B-Strict
    & 0.00/0.00
    & 0.37/0.28
    & 0.00/0.00
    & 0.00/0.00
    & 0.22/0.02
    & 0.12/0.06 \\

    \midrule[0.8pt]
        \textbf{Intrinsic Guard} \\
        Ling-3.0-flash-Vanilla & 0.00/0.00 & 0.28/0.03 & 0.00/0.00 & 0.00/0.00 & 0.12/0.00 & 0.08/0.01 \\
        Ling-3.0-flash-CC++ & 0.00/0.00 & 0.26/0.05 & 0.00/0.00 & 0.00/0.00 & 0.12/0.02 & 0.08/0.01 \\
        Ling-3.0-flash-Kelp & 0.00/0.00 & 0.46/0.07 & 0.00/0.00 & 0.00/0.00 & 0.96/0.00 & 0.28/0.01\\

  \midrule[0.8pt]

    \textbf{Ling-3.0-tiny-singprobe} & 0.00/0.00
      & 0.41/0.14
      & 0.00/0.00
      & 0.00/0.00
      & 0.87/0.05
      & 0.26/0.04 \\
        \textbf{Ling-3.0-flash-singprobe} & 0.00/0.00 & 0.25/0.07 & 0.00/0.00 & 0.00/0.00 & 0.13/0.00 & 0.08/0.01 \\






  \bottomrule
  \end{tabular}
  }
\end{table}

\partitle{False-positive Evaluation} We additionally evaluate false alarms under free generation using five nominally benign datasets. These datasets are MATH~\citep{hendrycks2021math}, databricks-dolly-15k~\citep{DatabricksBlog2023DollyV2}, GSM8K~\citep{cobbe2021gsm8k}, MBPP~\citep{austin2021mbpp}, and OpenBookQA~\citep{OpenBookQA2018}. Each model receives a query and generates its own response, during which we collect both query-intent and response-safety scores; the final predictions use the same aggregation rules described above. Table~\ref{tab:fpr} reports query/response FPR in each cell and macro averages across the five datasets. \guard maintains a low macro-average response FPR, with 0.04\% on Ling-3.0-tiny and 0.01\% on Ling-3.0-flash, on par with the strongest Qwen3Guard-Stream baselines. The matched intrinsic baselines also retain a 0.01\% macro-average response FPR; their query FPRs are 0.08\% for Vanilla and CC++, and 0.28\% for Kelp, compared with 0.08\% for \guard with Ling-3.0-flash. We note that these datasets are not exhaustively safety-audited and may contain a small number of genuinely unsafe queries or responses. The reported values should therefore be interpreted as apparent FPRs and may slightly overestimate the true false-positive rates.

\subsection{Response Hallucination Evaluation}

\begin{table}[t]
\centering
\small
\caption{Hallucination detection --- \textbf{response-level AUC} across 6 offline hallucination benchmarks. For each response, the Guard output is MAX-pooled across the response token mask, then the ROC AUC is computed sample-vs-sample (threshold-independent). \textbf{Macro} is the equal-weight mean of the per-benchmark AUCs.}
\label{tab:hallu_eval}
\resizebox{0.95\linewidth}{!}{%
\begin{tabular}{lcccccc>{\columncolor{green!8}}c}
\toprule
\textbf{Model} & \textbf{FactCHD} & \textbf{FaithDial} & \textbf{FAVA} & \textbf{RAGTruth} & \textbf{Shroom} & \textbf{WikiBio} & \textbf{Avg.} \\
\midrule
\textbf{Ling-3.0-tiny} \\
\addlinespace[2pt]
DRIFT & 0.7566 & 0.9030 & \textbf{0.5871} & 0.6771 & 0.6736 & 0.8477 & 0.7408 \\
HaMI & 0.6722 & 0.8230 & 0.5256 & 0.6168 & 0.6675 & 0.8381 & 0.6905 \\
SAPLMA-raw & 0.7389 & 0.8381 & 0.5516 & 0.6375 & 0.6460 & 0.6294 & 0.6736 \\
SAPLMA-response & 0.7018 & 0.8480 & 0.5756 & \textbf{0.6984} & 0.6832 & 0.9133 & 0.7367 \\

\textbf{Ling-3.0-tiny-singprobe} & \textbf{0.7819} & \textbf{0.9529} & 0.5802 & 0.6980 & 0.6946 & \textbf{0.9515} & \textbf{0.7765} \\
\midrule[1pt]
\textbf{Ling-3.0-flash} \\
\addlinespace[2pt]
DRIFT & 0.7972 & 0.9115 & \textbf{0.6577} & \textbf{0.7657} & 0.7266 & 0.9414 & 0.8000 \\
HaMI & 0.7571 & 0.8542 & 0.5801 & 0.6253 & 0.6899 & 0.9200 & 0.7378 \\
SAPLMA-raw & 0.7904 & 0.8601 & 0.5112 & 0.6845 & 0.6743 & 0.7828 & 0.7172 \\
SAPLMA-response & 0.7542 & 0.8776 & 0.5835 & 0.7538 & 0.7050 & \textbf{0.9701} & 0.7740 \\

\textbf{Ling-3.0-flash-singprobe} & \textbf{0.8308} & \textbf{0.9498} & 0.5996 & 0.7282 & \textbf{0.7314} & 0.9674 & \textbf{0.8012} \\
\bottomrule
\end{tabular}
}
\end{table}

We evaluate response-level hallucination detection on six offline benchmarks (FactCHD~\citep{chen2023factchd}, FaithDial~\citep{dziri2022faithdial}, FAVA~\citep{mishra2024fava}, RAGTruth~\citep{niu2024ragtruth}, Shroom~\citep{mickus2024semeval}, and WikiBio~\citep{manakul2023selfcheckgpt}). Given a query-response pair, \guard produces a hallucination score at every response token under teacher forcing. We use the maximum score over all response positions as the response-level risk score and compute ROC AUC across samples. Table~\ref{tab:hallu_eval} reports the per-benchmark results and their macro average.

\guard achieves strong and consistent performance across both base models. On Ling-3.0-tiny, \guard obtains the best average AUC of 0.7765, outperforming the strongest baseline, DRIFT, by 3.57 points, and achieves the best per-benchmark AUC on FactCHD (0.7819), FaithDial (0.9529), and WikiBio (0.9515). Scaling to Ling-3.0-flash further improves \guard to 0.8012, outperforming the best baseline DRIFT (0.8000). These results indicate that the hidden states of the base model provide effective signals for hallucination detection and that their quality improves with model scale, enabling a lightweight intrinsic head to be competitive with or surpass specialized hallucination detectors.

\subsection{Overhead Evaluation}

\begin{table}[t]
\centering
\caption{Relative latency overhead of \guard under different concurrency levels, reported as mean $\pm$ standard deviation over 8 measurement rounds. TTFT denotes time to first token, and ITL denotes inter-token latency. \textit{Decode Probe} runs \guard only during decoding, while \textit{Prefill-enabled Probe} additionally runs \guard during the prefill stage. Positive values indicate additional overhead relative to the base model without \guard.}
\label{tab:latency_concurrency}
\resizebox{0.70\linewidth}{!}{
\begin{tabular}{c|cc|cc}
\toprule
\multirow{2}{*}{\textbf{Concurrency}}
& \multicolumn{2}{c|}{\textbf{Decode Probe}}
& \multicolumn{2}{c}{\textbf{Prefill-enabled Probe}} \\
\cmidrule(lr){2-3} \cmidrule(lr){4-5}
& \textbf{TTFT ($\Delta$\%)}
& \textbf{ITL ($\Delta$\%)}
& \textbf{TTFT ($\Delta$\%)}
& \textbf{ITL ($\Delta$\%)} \\
\midrule
1  & $+0.68 \pm 0.22$ & $+0.18 \pm 0.15$ & $+0.89 \pm 0.29$ & $+0.18 \pm 0.15$ \\
2  & $-1.45 \pm 4.17$ & $+0.33 \pm 0.12$ & $+2.25 \pm 4.86$ & $+0.33 \pm 0.12$ \\
4  & $-1.91 \pm 5.97$ & $+0.30 \pm 0.12$ & $+1.85 \pm 7.77$ & $+0.38 \pm 0.10$ \\
8  & $+0.39 \pm 1.84$ & $+0.26 \pm 0.09$ & $+1.01 \pm 0.53$ & $+0.21 \pm 0.12$ \\
16 & $+0.06 \pm 3.57$ & $+0.42 \pm 0.09$ & $+0.86 \pm 1.15$ & $+0.39 \pm 0.12$ \\
32 & $+1.14 \pm 0.93$ & $+0.45 \pm 0.18$ & $+1.34 \pm 0.37$ & $+0.47 \pm 0.15$ \\
\bottomrule
\end{tabular}
}
\end{table}

We evaluate the inference overhead introduced by \guard (with the unified architecture on Ling-3.0-flash) under a controlled serving-load protocol. Each request uses an input length of 8{,}192 tokens and an output length of 1{,}024 tokens, and we vary the concurrency from 1 to 32. At each concurrency level, we submit a fixed number of prompts per round (24 for concurrency 1, 2, and 4; 48 for concurrency 8; and 96 for concurrency 16 and 32) and repeat the measurement over 8 independent rounds. Within each round, we compute the median TTFT and ITL, pair each \guard-enabled configuration with the base-model-only baseline of the same round, and report the mean and standard deviation of the paired relative differences across rounds. We consider two deployment modes of \guard: \textit{Decode Probe}, which runs the probe only during autoregressive decoding, and \textit{Prefill-enabled Probe}, which additionally invokes the probe during the prefill stage.

As shown in Table~\ref{tab:latency_concurrency}, the additional latency introduced by \guard remains small across all concurrency levels. For the \textit{Decode Probe} configuration, the ITL overhead is consistently below $0.5\%$, ranging from $+0.18\%$ at concurrency 1 to $+0.45\%$ at concurrency 32. The TTFT differences are within measurement noise: they are typically within $\pm 1.5\%$ and even show slightly negative point estimates at concurrency 2 and 4, with standard deviations that comfortably span zero. This is expected because the decode-only probe does not participate in the prefill phase and therefore should not change TTFT in a systematic way. Enabling the probe during prefill (\textit{Prefill-enabled Probe}) adds a small but consistently positive TTFT overhead, ranging from $+0.86\%$ to $+2.25\%$, while the ITL overhead remains essentially unchanged and below $0.5\%$ ($+0.18\%$ to $+0.47\%$). Overall, the relative overhead does not grow with serving concurrency, and even in the more expensive prefill-enabled setting the additional cost stays around $1$--$2\%$ on TTFT and well under $0.5\%$ on ITL. These results demonstrate that continuously extracting guardrail signals from the base model's hidden states introduces little additional cost to either first-token latency or token-by-token decoding, and that this overhead remains stable under increasing serving load. This efficiency supports the practical deployment of \guard as an intrinsic streaming guardrail without requiring a separate guardrail inference service.



\subsection{Ablaton Study of the Number of Tapped Layers}

\begin{figure}[t]
    \centering
    \includegraphics[width=0.99\linewidth]{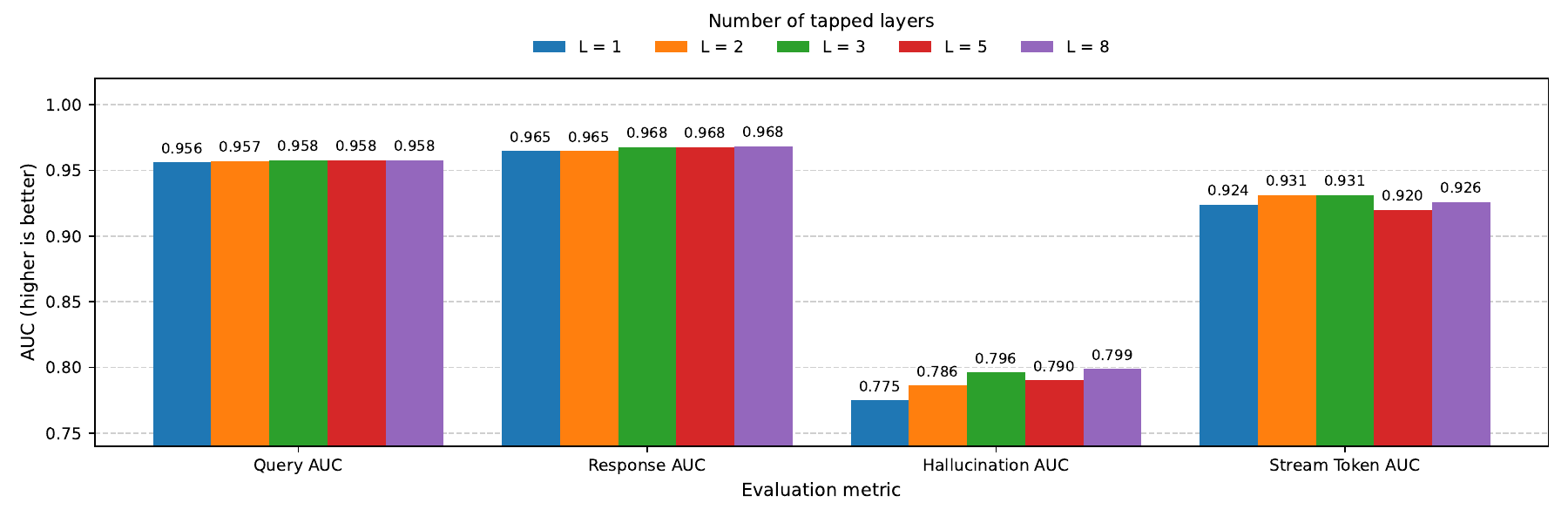}
    \vspace{-1em}
    \caption{The ablation study of the number of tapped base-model layers.}
    \label{fig:layer-ablation}
\end{figure}

We ablate the number of hidden layers tapped from the frozen base model, varying $L\in\{1,2,3,5,8\}$ while keeping the Guard architecture and training recipe fixed. As shown in Figure~\ref{fig:layer-ablation}, safety detection is generally robust to the number of tapped layers. Query AUC remains within a narrow range of $0.9560$-$0.9579$, while Response AUC varies from $0.9648$ to $0.9684$. Notably, using only the final layer already yields strong performance, achieving a Response AUC of $0.9648$ and a streaming response AUC of $0.9767$. Incorporating intermediate-layer representations nevertheless provides modest improvements: compared with $L{=}1$, the $L{=}3$ configuration improves Response AUC from $0.9648$ to $0.9677$ and streaming response AUC from $0.9767$ to $0.9823$. Hallucination detection benefits more clearly from multi-layer features, with AUC increasing from $0.7750$ at $L{=}1$ to $0.7963$ at $L{=}3$ and reaching $0.7988$ at $L{=}8$. This suggests that hallucination detection may require information that is less completely captured by the final-layer representation. Across all configurations, the benign false-positive rate remains at or below $0.02\%$, indicating that tapping additional layers does not noticeably increase over-triggering on benign inputs. However, tapping more layers increases the input dimensionality of the Guard, leading to a larger parameter count, higher memory consumption, and additional inference latency. Since the performance gains become marginal beyond $L{=}3$ while these deployment costs continue to grow, we use three tapped layers by default to achieve a favorable effectiveness--efficiency trade-off.

\subsection{Generation-time Applications of \guard}

\subsubsection{Application 1: \guard for Online Content Detection}

The preceding evaluations are conducted offline: each detector receives a fixed query--response pair under teacher forcing, allowing its predictions to be compared against pre-existing labels. In practical deployment, however, the base model generates responses autoregressively, and the guardrail must assess content produced by the model itself as generation unfolds. We therefore conduct an online evaluation in which Ling-3.0 freely generates responses while \guard produces detection signals alongside decoding. We study this setting for both response safety and hallucination detection.

\partitle{Online Safety Detection} We compare \guard with GPT-OSS-120B~\citep{agarwal2025gptoss} (as an LLM judger), Qwen3Guard-Stream-8B, and YuFeng-XGuard-Reason-8B on HarmBench, WildGuard, and XSTest. Because the safety labels of freely generated responses are not available a priori, we adopt a leave-one-out evaluation protocol. For each evaluated detector, the majority vote of the other three detectors is used as its pseudo-ground-truth label, ensuring that the evaluated detector does not contribute to its own reference label. We then report Accuracy and F1 against this leave-one-out consensus.

As shown in Table~\ref{tab:online_safety}, \guard performance improves substantially with a stronger base model. With Ling-3.0-tiny, \guard obtains an average Accuracy of 0.8089 and an average F1 of 0.4726, which trails the stronger standalone baselines under this leave-one-out protocol. With Ling-3.0-flash, however, \guard delivers the strongest overall performance, reaching an average Accuracy of 0.9641 and an average F1 of 0.6452. It outperforms the strongest baseline average F1 (YuFeng-XGuard-Reason-8B at 0.5964) by 4.88 points and obtains the best F1 on both WildGuard (0.6364) and XSTest (0.5405). These results demonstrate that \guard remains effective when its predictions are produced during free autoregressive generation, rather than from fixed responses under teacher forcing.

Note that the comparatively modest F1 scores are largely attributable to the severe class imbalance in this online setting: most freely generated responses are Safe, leaving relatively few Unsafe examples. Under such a low positive-class prevalence, even a small number of false positives or disagreements among the detectors can substantially reduce precision and, consequently, the F1 score. The online F1 values should therefore be interpreted together with Accuracy and primarily used for relative comparison among methods evaluated under the same protocol, rather than directly compared with results on more balanced offline benchmarks.

\begin{table}[t]
\centering
\caption{Online response safety detection under free autoregressive generation. For each evaluated detector, pseudo-ground-truth labels are obtained by majority vote of the other three detectors under a leave-one-out protocol. The \guard detector is evaluated with its bias-calibrated checkpoint at the canonical 0.5 decision threshold. Values report Accuracy (Acc) and F1. The best result within each backbone is highlighted in bold.}
\label{tab:online_safety}
\begin{tabular}{l c c c c c c >{\columncolor{green!8}}c >{\columncolor{green!8}}c}
\toprule
& \multicolumn{2}{c}{Harmbench} & \multicolumn{2}{c}{WildGuard} & \multicolumn{2}{c}{XSTest} & \multicolumn{2}{>{\columncolor{green!8}}c}{Avg.} \\
\cmidrule(lr){2-3}\cmidrule(lr){4-5}\cmidrule(lr){6-7}\cmidrule(lr){8-9}
Method & Acc & F1 & Acc & F1 & Acc & F1 & Acc & F1 \\
\midrule
\textbf{Ling-3.0-tiny} \\
\addlinespace[2pt]
GPT-OSS-120B
& 0.7996
& 0.6691
& 0.9039
& 0.4857
& \textbf{0.9200}
& \textbf{0.4088}
& 0.8745
& 0.5212 \\

Qwen3Guard-Stream-8B
& 0.8068
& 0.6766
& \textbf{0.9226}
& 0.5959
& 0.9193
& 0.2800
& \textbf{0.8829}
& 0.5175 \\

YuFeng-XGuard-Reason-8B
& \textbf{0.8195}
& \textbf{0.7358}
& 0.9141
& \textbf{0.6272}
& 0.9021
& 0.3907
& 0.8786
& \textbf{0.5846} \\

\textbf{Ling-3.0-tiny-singprobe}
& 0.7375
& 0.6600
& 0.8506
& 0.4875
& 0.8386
& 0.2703
& 0.8089
& 0.4726 \\

\midrule
\textbf{Ling-3.0-flash} \\
\addlinespace[2pt]
GPT-OSS-120B
& \textbf{0.9513}
& \textbf{0.8112}
& 0.9678
& 0.5630
& 0.9761
& 0.3846
& 0.9651
& 0.5863 \\

Qwen3Guard-Stream-8B
& 0.9358
& 0.7315
& \textbf{0.9755}
& 0.5782
& \textbf{0.9873}
& 0.2609
& \textbf{0.9662}
& 0.5235 \\

YuFeng-XGuard-Reason-8B
& 0.9391
& 0.7925
& 0.9661
& 0.5885
& 0.9783
& 0.4082
& 0.9612
& 0.5964 \\

\textbf{Ling-3.0-flash-singprobe}
& 0.9302
& 0.7586
& 0.9747
& \textbf{0.6364}
& \textbf{0.9873}
& \textbf{0.5405}
& 0.9641
& \textbf{0.6452} \\
\bottomrule
\end{tabular}
\end{table}

\begin{table}[t]
\centering
\caption{Online response hallucination detection under free autoregressive generation. Generated responses are assigned binary hallucination labels by Qwen3.5-397B-A17B based on the question, generated response, and reference answer. Values report response-level ROC AUC. The best result within each backbone is highlighted in bold.}
\label{tab:hallu_online_auc}
\resizebox{\textwidth}{!}{%
    \renewcommand{\arraystretch}{1.12}
    \setlength{\tabcolsep}{5.0pt}
\begin{tabular}{lccccccc >{\columncolor{green!8}}c}
\toprule
Method
    & \textbf{AA-Omniscience}
    & \textbf{BBH}
    & \textbf{GSM8K}
    & \textbf{MATH-500}
    & \textbf{NQ-Open}
    & \textbf{SQuAD}
    & \textbf{TruthfulQA}
    & Avg. \\
\midrule

\multicolumn{9}{l}{\textbf{Ling-3.0-tiny}} \\
\addlinespace[2pt]
DRIFT
    & 0.6380 & 0.6058 & 0.6446 & 0.4437
    & 0.7482 & \textbf{0.6180} & \textbf{0.7229} & 0.6316 \\
HaMI
    & 0.5643 & 0.6431 & 0.5386 & 0.4185
    & 0.6603 & 0.5570 & 0.5859 & 0.5668 \\
SAPLMA-raw
    & 0.5928 & 0.5924 & 0.6368 & 0.5448
    & 0.5882 & 0.5616 & 0.5661 & 0.5832 \\
SAPLMA-response
    & \textbf{0.6695} & \textbf{0.6767} & 0.5426 & 0.4348
    & 0.7089 & 0.6076 & 0.6411 & 0.6116 \\
\textbf{Ling-3.0-tiny-singprobe}
    & 0.6032
    & 0.5891
    & 0.7328
    & \textbf{0.7292}
    & \textbf{0.8136}
    & 0.6009
    & 0.6813
    & \textbf{0.6786} \\

\addlinespace[3pt]
\cmidrule(lr){1-9}
\addlinespace[2pt]

\multicolumn{9}{l}{\textbf{Ling-3.0-flash}} \\
\addlinespace[2pt]
DRIFT
    & 0.6662 & 0.5808 & 0.6193 & 0.6348
    & 0.6852 & 0.7420 & 0.6536 & 0.6546 \\
HaMI
    & 0.5312 & 0.5802 & 0.6295 & 0.5519
    & 0.6888 & 0.7143 & 0.6389 & 0.6193 \\
SAPLMA-raw
    & 0.6799 & 0.4872 & 0.6113 & 0.4877
    & 0.6062 & 0.7339 & 0.5879 & 0.5992 \\
SAPLMA-response
    & 0.5188 & \textbf{0.5843} & 0.6495 & 0.6079
    & 0.7125 & 0.7247 & \textbf{0.6699} & 0.6382 \\
\textbf{Ling-3.0-flash-singprobe}
    & 0.7353
    & 0.4884
    & \textbf{0.7702}
    & \textbf{0.7900}
    & \textbf{0.7792}
    & \textbf{0.8597}
    & 0.6668
    & \textbf{0.7271} \\

\bottomrule
\end{tabular}%
}
\end{table}

\partitle{Online Hallucination Detection} In the online hallucination evaluation, the model first performs autoregressive rollout on each test set to generate responses. An LLM, specifically Qwen3.5-397B-A17B in our experiments, then judges whether each response is correct based on the question, the model response, and the ground-truth answer, producing a binary label of whether hallucination occurs. Using these labels, we compute the hallucination-detection AUC for five methods: DRIFT~\citep{bhatnagar2026drift}, HaMI~\citep{niu2025robust}, SAPLMA-raw, SAPLMA-response~\citep{azaria2023internal}, and \guard, across seven datasets: AA-Omniscience~\citep{jackson2025aaomniscience}, BBH~\citep{suzgun2023bbh}, GSM8K~\citep{cobbe2021gsm8k}, MATH-500~\citep{hendrycks2021math}, NQ-Open~\citep{lee2019nqopen}, SQuAD~\citep{rajpurkar2016squad}, and TruthfulQA~\citep{lin2022truthfulqa}. Results are reported on two backbones, Ling-3.0-tiny and Ling-3.0-flash.


The results show that \guard achieves a clear and consistent advantage on both backbones, attaining the best average AUC in every case. On Ling-3.0-tiny, it reaches 0.6786, leading the second-best method DRIFT (0.6316) by an absolute margin of 0.047. On Ling-3.0-flash, it reaches 0.7271, leading DRIFT (0.6546) by 0.073. As backbone capability increases, \guard's performance improves from 0.6786 to 0.7271, and its lead over the second-best method widens further, indicating that \guard can make fuller use of the internal signals from stronger backbone models.

\guard's advantage is also evident on individual datasets. On Ling-3.0-flash, \guard ranks first on five of the seven datasets: AA-Omniscience, GSM8K, MATH-500, NQ-Open, and SQuAD. Among them, the AUC on four datasets, AA-Omniscience, GSM8K, MATH-500, and SQuAD, exceeds 0.73, with SQuAD reaching 0.8597. On Ling-3.0-tiny, \guard also leads comprehensively on knowledge- or reasoning-intensive tasks such as GSM8K, MATH-500, and NQ-Open, achieving 0.7328, 0.7292, and 0.8136, respectively. In contrast, the baselines achieve local best results only on a few individual datasets and do not show a consistent advantage across backbones. Although SAPLMA-response performs relatively well on some BBH and TruthfulQA settings, its average AUC is 0.067 lower than \guard on tiny and 0.089 lower on flash; the gaps for HaMI and SAPLMA-raw are even larger.

\subsubsection{Application 2: \guard for Constrained Safe Decoding}

\begin{figure*}[t]
\centering
\begin{subfigure}[t]{0.48\linewidth}
\centering
\includegraphics[width=\linewidth]{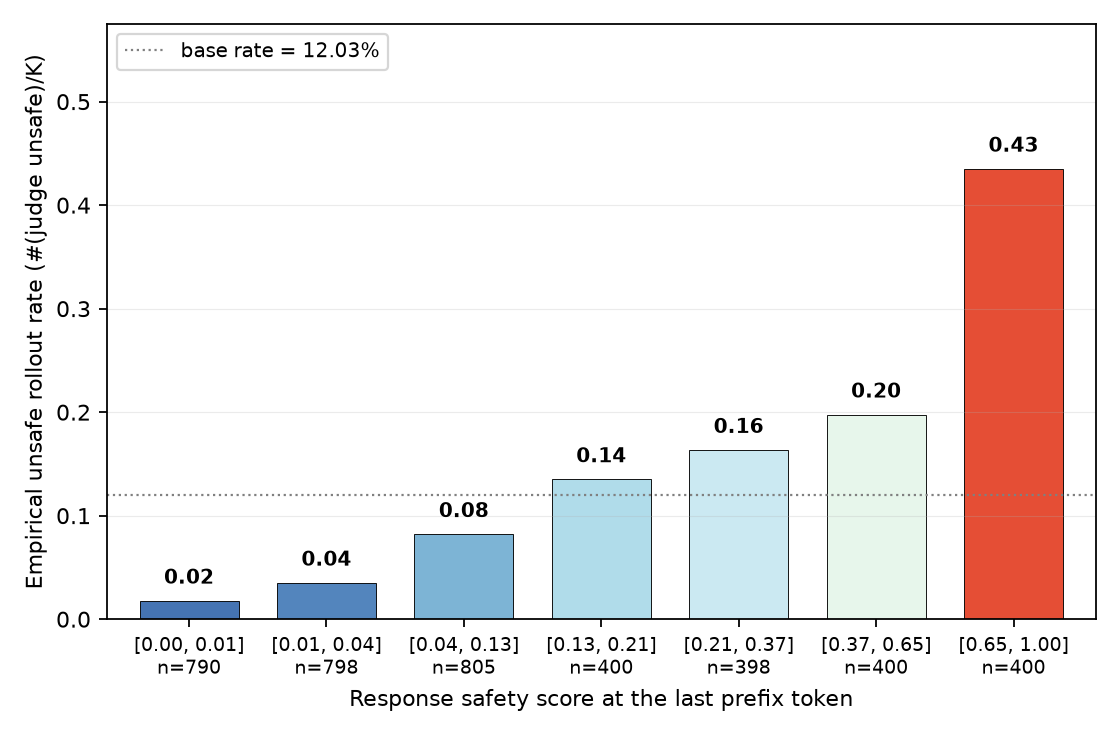}
\caption{Predicting future generation risk from annotated safe prefixes.
Starting from each query and its annotated safe response prefix, we group subsequent rollouts by the \guard score at the final safe-prefix token and report the empirical unsafe rates of their rollouts. Higher safe-prefix scores consistently correspond to greater unsafe risk in future continuations.}
\label{fig:rollout}
\end{subfigure}
\hfill
\begin{subfigure}[t]{0.48\linewidth}
\centering
\includegraphics[width=\linewidth]{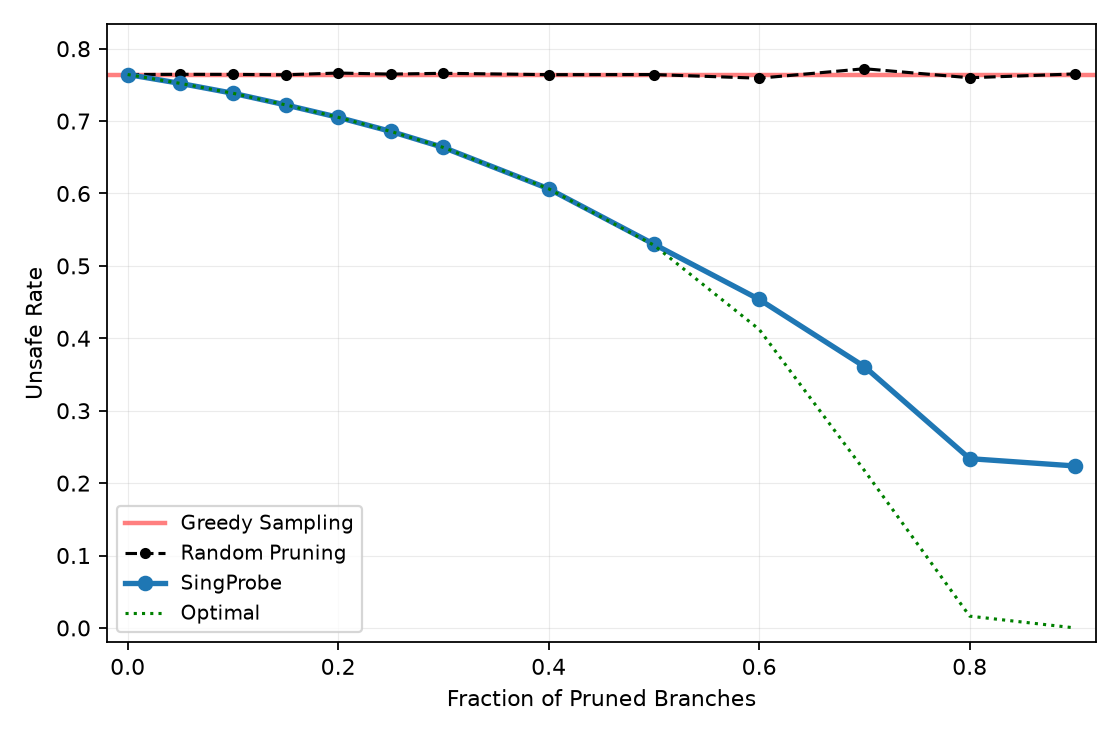}
\caption{\guard-guided branch pruning from annotated safe prefixes.
Starting from each query and its annotated safe response prefix, we sample intermediate chunks to form candidate decoding branches. Branches with the highest chunk-end \guard scores are pruned before further continuation. The resulting policy substantially reduces the unsafe rate compared with greedy sampling and random pruning and closely approaches the optimal.}
\label{fig:chunk-prune}
\end{subfigure}
\caption{Anticipatory safety prediction and constrained safe decoding with \guard.}
\label{fig:constrained_decoding}
\end{figure*}

Beyond detecting unsafe content that has already appeared, an intrinsic streaming guardrail may also reveal how the safety of a response is likely to evolve. We study this possibility through two sequential research questions. First, \emph{can \guard predict the safety tendency of future generations from the current prefix?} Second, if the score is predictive of future risk, \emph{can it be used to guide safe decoding by terminating risky branches before they continue?} Figure~\ref{fig:constrained_decoding} summarizes the corresponding rollout prediction and branch-pruning experiments.

\partitle{Experimental Settings} We use Ling-3.0-flash as the frozen generation model and its \guard head as the safety probe. The response-unsafety score is streamed directly from the hidden states produced during decoding through the same SGLang serving pipeline, without an additional forward pass. For every evaluated prefix, we sample eight continuations. An independent YuFeng-XGuard-Reason-8B judge assigns a binary Safe or Unsafe label to each completed rollout. It is worth noting that the safety of the generated responses is assessed by an external judge and therefore may not constitute ground-truth labels. Nevertheless, the observed increase or decrease in the unsafe response rate still provides meaningful evidence of \guard's contribution to safety.

\textbf{Can \guard predict future generation risk?} We operationalize this question by testing whether the score at the current \emph{safe} prefix correlates with the fraction of future continuations that eventually become unsafe. Specifically, we collect 500 query--prefix pairs, including 350 unsafe-origin and 150 safe-origin examples. Each input consists of a query followed by an annotated safe response prefix, allowing us to test whether \guard can anticipate a future transition to unsafe content while the observed response remains safe. We record \guard's response-unsafety score at the final token of this safe prefix, before any further continuation is sampled, and then generate eight independent rollouts. Their judged labels define the empirical future unsafe rate of the query--prefix pair, $\hat{p}_{\mathrm{unsafe}}=n_{\mathrm{unsafe}}/8$.

Figure~\ref{fig:rollout} shows a clear monotonic relationship between the safe-prefix score and subsequent rollout safety. The empirical unsafe rate rises from approximately $0.02$ in the lowest-score interval to $0.43$ in the highest, which is roughly $3.6\times$ the overall base rate of $0.1203$ and over $21\times$ the rate in the lowest-score group. The increase becomes particularly pronounced in the higher-score intervals, rising from $0.08$ to $0.14$, $0.16$, $0.20$, and finally $0.43$. Because every observed prefix is annotated as safe and the score is recorded before continuation, this trend cannot be explained by \guard merely reacting to harmful content that has already appeared. Instead, it indicates that the hidden state at a safe prefix contains information about the conditional safety tendency of its possible continuations.

\textbf{Can \guard guide safe decoding?} Motivated by this predictive relationship, we next test whether the score can identify risky decoding branches early enough to prune them. As in the first experiment, each root consists of a query followed by an annotated safe response prefix. We select 16 conflict query--prefix roots for which three to five of the initial eight rollouts are unsafe, ensuring that every root admits both safe and unsafe continuations. Starting from each query and safe prefix, we sample 24 distinct 64-token chunks, yielding 384 intermediate branch prefixes. At the end of each chunk, we record \guard's response-unsafety score before sampling any further tokens and then generate eight continuations, producing 3,072 judged rollouts. 
We simulate constrained decoding by removing the fraction $p$ of branches with the highest \guard unsafety scores and measuring the unsafe rate among continuations from the retained branches. As shown in Figure~\ref{fig:chunk-prune}, this \guard-guided policy is compared with random pruning averaged over 200 permutations and an empirical-risk oracle that removes branches with the highest observed unsafe rates.

Guard-guided pruning consistently reduces the unsafe rate of the retained rollout pool, whereas random pruning leaves it nearly unchanged at approximately $0.76$. When pruning $50\%$ of the branches, \guard lowers the retained-pool unsafe rate to $0.53$, a relative reduction of approximately $30\%$ compared with random pruning, and essentially matches the oracle rate of $0.53$. The unsafe rate continues to decrease as more high-scoring branches are removed, reaching approximately $0.22$ when $90\%$ of branches are pruned. The \guard and oracle curves nearly overlap up to a pruning fraction of about $0.5$ and diverge only at aggressive pruning ratios, where the oracle drops toward $0$ while \guard plateaus around $0.22$--$0.23$. This suggests that \guard is highly accurate at identifying the highest-risk branches but is less precise in ranking the remaining lower-risk branches. Overall, these results demonstrate that \guard's anticipatory signal is not only correlated with future safety outcomes but can also serve as a practical control signal for constrained safe decoding.

\section{\guard-Med: On-Demand Medical-Risk Intervention}
\label{sec:singprobe_med}

\subsection{Motivation and Framework Overview}

The preceding sections introduce \guard as a lightweight guardrail for a frozen LLM during autoregressive generation. We now study how such token-level internal-state signals can be used not only for detection, but also to control \emph{when} a generation-time intervention should modify subsequent decoding.

Medical generation provides a representative setting in which selective intervention is particularly important. A harmful recommendation may emerge only after a clinically sound prefix, making query-level screening insufficient and post-hoc review too late to prevent the error from being generated~\citep{singhal2023large,yang2024medguard,draelos2026unsafe, mishra2024finegrained}. At the same time, applying a corrective decoding mechanism throughout the entire response is undesirable: always-on steering can perturb otherwise correct medical content and incurs persistent decoding cost. An effective intervention system should therefore answer two complementary questions: \emph{when should intervention be activated, and how should generation be modified once intervention is needed?}

\begin{figure}[t]
    \centering
    \includegraphics[width=0.92\linewidth]{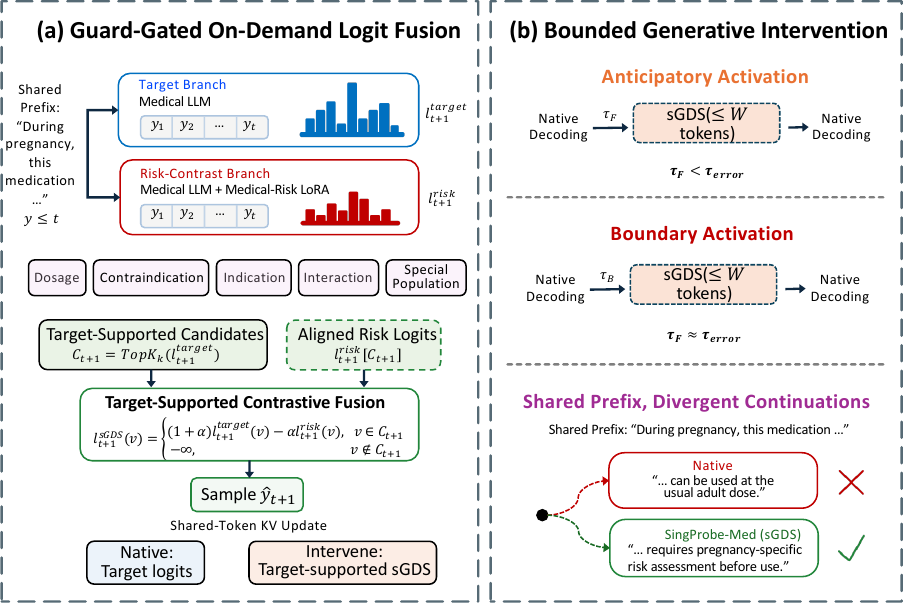}
    \caption{\textbf{SingProbe-Med intervention overview.} (a) Once admitted, target-supported sGDS contrasts target and medical risk-pattern logits while keeping both branches synchronized through a shared sampled token. (b) Future- or Boundary-based activation opens a bounded intervention window, after which generation returns to native decoding.}
    \label{fig:singprobemed-overview}
\end{figure}

\guard-Med separates these two responsibilities. As illustrated in Figure~\ref{fig:singprobemed-overview}, \guard-Med is responsible for deciding when to intervene by continuously monitoring the evolving generation trajectory through lightweight internal-state signals. Once the admission condition is satisfied, a target-supported intervention mechanism named support-constrained Guided Decoding Steering (sGDS) determines how to intervene by contrastively re-ranking the target model's candidate tokens over a localized risky suffix. The intervention is bounded in duration, after which generation returns to the target model's native decoding. This design separates inexpensive continuous monitoring from the more expensive corrective decoding mechanism, avoiding persistent dual-branch inference on generations for which no intervention is needed.

We instantiate this framework on AntAngelMed~\citep{AntAngelMed}, a 100B-parameter MoE medical LLM. We next describe the two components separately: the \guard-Med admission mechanism for determining \emph{when to intervene}, followed by target-supported sGDS for determining \emph{how to intervene}.

\subsection{\guard-Med: When to Intervene}

At decoding step $t$, \guard-Med derives four complementary signals from the current prefix $y_{\leq t}$. Rather than serving as independent prediction tasks, these signals correspond to different decisions required by the intervention controller. The semantics of the four signals are summarized in Table~\ref{tab:singprobemed-signals}.
\begin{itemize}[leftmargin=*]
    \item \textbf{Intervention Eligibility $E_t$} determines whether the current prefix contains a clinically consequential proposition for which medical-risk intervention is applicable.
    \item \textbf{Global Risk $G_t$} estimates whether the evolving response trajectory is moving toward a harmful medical conclusion. The remaining two signals provide more localized timing information.
    \item \textbf{Future Risk $F_t$} indicates whether a confirmable medical error is likely to emerge within the next 1--32 tokens, enabling intervention before the error is committed.
    \item \textbf{Error Boundary $B_t$} identifies the first prefix at which a medical error becomes confirmable, allowing immediate containment when sufficiently early predictive evidence is unavailable.
\end{itemize}

\begin{table}[t]
    \centering
    \small
    \caption{Signal semantics in the SingProbe-Med intervention policy.}
    \label{tab:singprobemed-signals}
    \begin{tabular}{@{}
        >{\raggedright\arraybackslash}p{0.18\linewidth}
        >{\raggedright\arraybackslash}p{0.24\linewidth}
        >{\raggedright\arraybackslash}p{0.50\linewidth}@{}}
        \toprule
        Signal & Scope & Intervention semantics \\
        \midrule
        Intervention Eligibility $E_t$ & Intervention applicability & Determines whether the current prefix expresses a clinically consequential proposition for which medical-risk intervention is applicable. \\
        Global Risk $G_t$ & Response trajectory & Estimates whether the evolving response is moving toward a harmful medical conclusion. \\
        Future Risk $F_t$ & Pre-error timing & Indicates that a confirmable medical error is likely within the next 1--32 tokens, allowing intervention before the error is committed. \\
        Error Boundary $B_t$ & Error-onset timing & Marks the first prefix at which a medical error becomes confirmable, allowing immediate containment when advance evidence is unavailable. \\
        \bottomrule
    \end{tabular}
\end{table}

These signals are combined through a fixed evidential hierarchy to form an admission policy. Intervention of the next uncommitted token is activated according to
\begin{equation}
a_{t+1}
=
\mathbb{I}
\left[
E_t \geq \gamma_E
\;\wedge\;
\left(
G_t \geq \gamma_G
\;\vee\;
\left(
G_t \geq \gamma_C
\wedge
\left(
F_t \geq \gamma_F
\vee
B_t \geq \gamma_B
\right)
\right)
\right)
\right].
\label{eq:med_admission}
\end{equation}

This rule follows an $E \rightarrow G \rightarrow (F/B)$ hierarchy motivated by the asymmetric cost of false intervention. $E_t$ first restricts intervention to prefixes for which risk-directed steering is applicable. $G_t$ then provides trajectory-level evidence of emerging medical harm. When the global-risk signal is sufficiently strong, intervention can be admitted directly to avoid unnecessary delay. Under intermediate global evidence, the shorter-horizon $F_t$ and $B_t$ signals refine the intervention timing.

Importantly, $F_t$ and $B_t$ are complementary entry signals rather than sequential stages. $F_t$ supports \emph{anticipatory intervention} before a medical error becomes explicit, whereas $B_t$ provides \emph{boundary-level containment} at the earliest prefix where the error can be confirmed. All four signals are computed in parallel; the hierarchy therefore specifies evidential precedence rather than a sequential execution pipeline. The operating thresholds and release conditions are fixed on the development split before final evaluation.




\subsection{Target-Supported sGDS: How to Intervene}


Once \guard-Med determines that intervention is warranted, we steer the subsequent generation by explicitly modeling and suppressing medical-risk preferences in the decoding distribution. Specifically, we construct a risk-pattern branch by adapting the target model with a medical-risk LoRA adapter~\citep{hu2022lora}, and use the difference between the target and risk-pattern logits to guide contrastive intervention. In the following sections, we first define the medical-risk patterns covered by the intervention. We then describe how the corresponding risk-pattern branch is constructed and how its logits are incorporated into target-supported sGDS.

\partitle{Medical-risk Intervention Scope}
We define the intervention scope using five operational medical-risk categories: dosage and administration errors, special-population contraindications, wrong indications, dermatology-specific wrong indications, and dermatology-specific special-population contraindications. These categories are used to organize the risk-pattern adaptation data and jointly construct the training set for the medical intervention branch. The sampler maintains approximately balanced coverage across the five categories. Table~\ref{tab:singprobemed-risk-categories} summarizes the corresponding intervention targets. The resulting risk-pattern dataset is used to adapt the medical intervention branch described next.

\begin{table}[t]
    \centering
    \small
    \caption{Medical-risk categories covered by the intervention branch.}
    \label{tab:singprobemed-risk-categories}
    \begin{tabular}{@{}
        >{\raggedright\arraybackslash}p{0.31\linewidth}
        >{\raggedright\arraybackslash}p{0.61\linewidth}@{}}
        \toprule
        Category & Intervention target \\
        \midrule
        Dosage and administration
        & Incorrect dose, frequency, duration, or usage instruction. \\
        Special-population contraindication
        & A recommendation that conflicts with constraints associated with a
        clinically relevant patient population. \\
        Wrong indication
        & A medication or treatment recommendation that does not match the
        stated condition. \\
        Dermatology wrong indication
        & An indication mismatch involving a dermatologic condition or therapy. \\
        Dermatology special-population contraindication
        & A dermatologic recommendation that conflicts with population-specific
        safety constraints. \\
        \bottomrule
    \end{tabular}
\end{table}

\partitle{Medical Risk-pattern Branch}
We construct the medical risk-pattern branch by adapting the deployed AntAngelMed model with a LoRA module on the risk-pattern dataset described above. The adaptation follows a standard supervised fine-tuning style objective: given a medical prompt $x$ and a risk-pattern response $y=(y_1,\ldots,y_T)$, we optimize only the LoRA parameters $\Delta_{\mathrm{risk}}$ while keeping the base-model parameters $\theta$ frozen. Specifically, the branch is trained with the autoregressive language-modeling loss
\begin{equation}
\mathcal{L}_{\mathrm{risk}}
=
-\sum_{t=1}^{T}
\log
p_{\theta+\Delta_{\mathrm{risk}}}
\left(
y_t
\mid
x,y_{<t}
\right),
\label{eq:singprobemed-risk-training}
\end{equation}
where the training responses are drawn from the five medical-risk categories defined above. Optimizing this objective increases the adapted branch's likelihood on continuations that exhibit the corresponding medical-risk patterns, thereby encoding these patterns into the learned LoRA parameters.

After adaptation, the original model and the risk-pattern-adapted model form the two branches used during intervention. Let $f_{\theta}$ denote the original AntAngelMed model and $\Delta_{\mathrm{risk}}$ the learned medical-risk LoRA parameters. Given the realized decoding context $c_t=(x,y_{\leq t})$, they produce next-token logits as
\begin{equation}
\ell^{\mathrm{tar}}_{t+1}
=
f_{\theta}(c_t),
\qquad
\ell^{\mathrm{risk}}_{t+1}
=
f_{\theta+\Delta_{\mathrm{risk}}}(c_t).
\label{eq:singprobemed-branches}
\end{equation}

Because the adapted branch is explicitly trained on risk-pattern continuations, it assigns relatively greater support to tokens and continuations that are compatible with the covered medical-risk patterns. We therefore use it as an \emph{anti-expert}: $\ell^{\mathrm{risk}}_{t+1}$ provides a token-level risk preference that can be contrasted with the target-model logits during intervention, rather than being interpreted as an explicit medical-error probability or an independently generated correction.

\partitle{On-demand Target-supported sGDS}
With the medical risk-pattern branch defined above, we use its next-token logits as an anti-expert signal to modify the target model only when intervention is activated. At each activated decoding step, the target and risk-pattern branches evaluate the same realized context $c_t=(x,y_{\leq t})$, producing $\ell^{\mathrm{tar}}_{t+1}$ and $\ell^{\mathrm{risk}}_{t+1}$ for the same uncommitted token. We then contrast these two distributions while restricting the intervention to candidates already supported by the target model. Specifically, the target branch first defines the admissible candidate set
\begin{equation}
C_{t+1}=\operatorname{TopK}\left(\ell^{\mathrm{tar}}_{t+1},k\right).
\label{eq:singprobemed-support}
\end{equation}
Within this support, sGDS computes
\begin{equation}
\ell^{\mathrm{sGDS}}_{t+1}(v)=
\begin{cases}
(1+\alpha)\ell^{\mathrm{tar}}_{t+1}(v)-\alpha\ell^{\mathrm{risk}}_{t+1}(v), & v\in C_{t+1},\\
-\infty, & v\notin C_{t+1}.
\end{cases}
\label{eq:singprobemed-sgds}
\end{equation}
The resulting $\ell^{\mathrm{sGDS}}_{t+1}$ serves as the final next-token logits, from which the sampling distribution is obtained and the next token is selected. Equivalently, sGDS adds the contrastive direction $\alpha(\ell^{\mathrm{tar}}_{t+1}-\ell^{\mathrm{risk}}_{t+1})$ to the native target logits within $C_{t+1}$. Since the risk-pattern branch is adapted to favor continuations exhibiting the covered medical-risk patterns, candidates that receive relatively stronger support from this branch are suppressed, while candidates favored more strongly by the target branch are promoted relative to their competitors. The top-$k$ restriction keeps the target model as the source of admissible next tokens, so the risk branch only re-ranks plausible target-supported candidates rather than introducing its own off-support continuations.

The two branches operate as paired conditional scorers rather than independently decoded models. They begin from the same prompt and realized response prefix while maintaining separate model states and KV caches. At each activated step, a single token $\hat{y}_{t+1}$ is sampled from the fused sGDS distribution and appended to both branches. This shared-token update keeps their prefixes synchronized, ensuring that subsequent target and risk-pattern logits continue to describe the same generation history. If the branches were decoded independently, their prefixes would diverge and the resulting logit difference would no longer isolate the learned risk preference.

When $a_{t+1}=0$, generation proceeds directly from the unmodified target-model logits and the risk-pattern branch does not participate in logit fusion. Once intervention is activated, it opens an intervention episode
\begin{equation}
e=(\tau,W,k,\alpha),
\label{eq:singprobemed-episode}
\end{equation}
where $\tau$ denotes the activation position, $W$ bounds the intervention duration, $k$ determines the target-supported candidate set, and $\alpha$ controls the strength of the contrastive correction. We evaluate both a full intervention and a bounded variant. The full intervention uses $k=50$ and $\alpha=2$ until release, whereas the bounded variant uses $k=100$, $\alpha=3$, and $W=64$. After the episode is released, decoding returns to target-only logits on the already modified prefix.

Thus, the intervention is on demand at two levels. SingProbe-Med determines \emph{when} the more expensive dual-branch decoding should be activated, while target-supported sGDS determines \emph{how} the next-token distribution should be modified once activated. By confining both the activation interval and the admissible vocabulary, the resulting intervention suppresses learned medical-risk preferences only over the risk-relevant portion of generation.

\subsection{Evaluation}
\label{sec:singprobemed-eval}
We evaluate \guard-Med along four complementary quantitative axes: admission selectivity and timing, medical error correction, general capability preservation, and inference efficiency. We further provide a qualitative case study to illustrate how the resulting intervention modifies representative medical-risk continuations. All admission thresholds and episode-release conditions are selected on the development split and frozen before evaluation on the held-out medical set. No threshold is reselected on the test set or separately for an individual capability benchmark.

\subsubsection{Admission Selectivity and Timing}

We first evaluate whether the admission mechanism can identify trajectories that warrant intervention and trigger sufficiently early. ``Harm'' denotes a trajectory containing a target medical risk, whereas ``White'' denotes a trajectory for which intervention should not be triggered. Let $A_i$ indicate whether trajectory $i$ opens at least one intervention episode. At a fixed operating point, we report
\begin{equation}
\left\{
\begin{aligned}
\mathrm{White\ FPR}
&= \Pr(A_i=1\mid i\in\mathrm{White}), \\
\mathrm{Harm\ recall}
&= \Pr(A_i=1\mid i\in\mathrm{Harm}), \\
\mathrm{Precision}
&= \Pr(i\in\mathrm{Harm}\mid A_i=1).
\end{aligned}
\right.
\label{eq:singprobemed-operating-metrics}
\end{equation}
We additionally report ROC--AUC, Average Precision (AP), and PR--AUC from the complete trajectory-level ranking on the held-out set. For the White-FPR operating points $q\in\{10\%,5\%,1\%\}$, thresholds are selected on the development set and frozen before evaluating Harm recall and precision on the held-out set.


For the combined detector, Intervention Eligibility acts as a binary gate on the Global Risk score. Specifically, we define the gated risk score as
\begin{equation}
S_{E+G}
=
\mathbb{I}(E>\gamma_E)\,\sigma(G),
\label{eq:singprobemed-gated-risk}
\end{equation}
where $\gamma_E$ is the Intervention Eligibility threshold and $\sigma(\cdot)$ denotes the sigmoid function. Thus, Global Risk contributes to the ranking only when the corresponding prefix is considered eligible for medical-risk intervention.

Table~\ref{tab:singprobemed-signal-results} compares Global Risk alone with this eligibility-gated Global Risk score. Applying the Intervention Eligibility gate improves ROC--AUC, AP, and PR--AUC by 0.025, 0.012, and 0.011, respectively, indicating that suppressing risk scores on intervention-ineligible prefixes improves trajectory selection. Tightening the allowed White FPR from 10\% to 1\% increases precision from 92.34\% to 96.77\%, while Harm recall decreases from 51.70\% to 11.98\%, exposing a direct trade-off between intervention coverage and false-trigger control.

\begin{table}[t]
    \centering
    \caption{Operating characteristics of the SingProbe-Med admission signals
    on the held-out medical set.}
    \label{tab:singprobemed-signal-results}
    \begingroup
    \small
    \renewcommand{\arraystretch}{1.08}

    \begin{minipage}[t]{0.60\linewidth}
        \textbf{(a) Ranking quality}
        \par\vspace{0.25em}
        \begin{tabularx}{\linewidth}{@{}Xccc@{}}
            \toprule
            Detector & ROC-AUC & AP & PR-AUC \\
            \midrule
            Global Risk
            & 0.828 & 0.903 & 0.903 \\
            Intervention Eligibility $+$ Global Risk
            & \textbf{0.853} & \textbf{0.915} & \textbf{0.914} \\
            \bottomrule
        \end{tabularx}
    \end{minipage}
    \hfill
    \begin{minipage}[t]{0.36\linewidth}
        \textbf{(b) Constrained operating points (\%)}
        \par\vspace{0.25em}
        \begin{tabularx}{\linewidth}{@{}Xcc@{}}
            \toprule
            White FPR & Harm recall & Precision \\
            \midrule
            $\leq10\%$ & 51.70 & 92.34 \\
            $\leq5\%$  & 32.44 & 93.93 \\
            $\leq1\%$  & 11.98 & 96.77 \\
            \bottomrule
        \end{tabularx}
    \end{minipage}

    \endgroup
\end{table}

We further evaluate activation timing on Harm trajectories with a confirmable medical error boundary. Let $T^{\mathrm{confirm}}$ denote the first prefix at which the medical error becomes confirmable. Among Harm trajectories, 54.99\% of first activations occur at or before $T^{\mathrm{confirm}}$, meaning that the policy opens a correction window no later than the confirmable error boundary for more than half of the risky trajectories. Evaluating all four \guard-Med signals over 3,044,056 token states takes 0.019749 ms/token, corresponding to approximately 50,635 token states per second.

\begin{table}[t]
    \centering
    \small
    \caption{Paired medical correction results (\%). ``Activated'' conditions on trajectories where SingProbe-Med opens an sGDS episode.}
    \label{tab:singprobemed-repair-results}
    \begin{tabular}{@{}lrrrr@{}}
        \toprule
        & \multicolumn{2}{c}{Full intervention} & \multicolumn{2}{c}{64-token window} \\
        \cmidrule(lr){2-3}\cmidrule(lr){4-5}
        Metric & Overall & Activated & Overall & Activated \\
        \midrule
        Broad mitigation (ERROR $\rightarrow$ non-ERROR) & \textbf{25.03} & \textbf{30.27} & 24.30 & 29.39 \\
        Strict repair (ERROR $\rightarrow$ SAFE) & \textbf{12.58} & \textbf{15.22} & 11.72 & 14.17 \\
        Safe regression & 1.84 & 6.78 & 2.53 & 9.32 \\
        \bottomrule
    \end{tabular}
    \vspace{-1em}
\end{table}

\subsubsection{Medical Error Correction}

We next evaluate whether an admitted intervention corrects the final medical outcome. Each test case is matched by prompt between target-only decoding and an intervention configuration. Let $z_i^0$ and $z_i^m$ denote the three-way adjudicated outcomes of the baseline and intervened responses in $\{\mathrm{ERROR},\mathrm{SAFE},\mathrm{UNCERTAIN}\}$. We define \emph{broad mitigation} as a baseline ERROR changed to either SAFE or UNCERTAIN, \emph{strict repair} as ERROR $\rightarrow$ SAFE, and \emph{safe regression} as a baseline SAFE changed to a non-SAFE outcome. ``Overall'' reports these transitions over the corresponding baseline population in the complete test set, whereas ``Activated'' additionally conditions on trajectories where at least one sGDS episode is opened. We compare two intervention configurations. The full intervention uses $k=50$ and $\alpha=2$ until release, whereas the bounded configuration uses $k=100$, $\alpha=3$, and a maximum intervention window of $W=64$ tokens. 

Table~\ref{tab:singprobemed-repair-results} reports the paired correction results. The full intervention mitigates 25.03\% of baseline erroneous generations over the complete test set, including 12.58\% that are strictly repaired from ERROR to SAFE. On trajectories where intervention is actually activated, broad mitigation and strict repair increase to 30.27\% and 15.22\%, respectively, indicating that the admission mechanism concentrates intervention on trajectories with greater correction potential. Restricting intervention to a 64-token window preserves most of this benefit. The bounded configuration achieves 24.30\% broad mitigation and 11.72\% strict repair overall, retaining 97.1\% and 93.2\% of the corresponding full-intervention results. This suggests that most useful corrections occur in the early suffix following activation rather than requiring persistent dual-branch decoding.

Intervention nevertheless introduces a non-zero risk of modifying originally safe generations. Reported safe regression is 1.84\% overall for the full intervention and 2.53\% for the 64-token configuration, with higher rates when conditioning specifically on activated trajectories. This trade-off further motivates selective admission rather than applying sGDS indiscriminately.

\subsubsection{General Capability Preservation}

We next examine whether selective intervention perturbs model capability on inputs for which medical-risk steering is unnecessary. We compare native AntAngelMed-100B, always-on sGDS, and \guard-Med across AIME24/25~\citep{maxwelljia2024aime, aime25}, HealthBench~\citep{arora2025healthbench}, CFBench~\citep{zhang2025cfbench}, Arena-Hard-v2~\citep{li2025arenahard}, AlignBench~\citep{liu2024alignbench}, and WritingBench~\citep{wu2026writingbench}. Evaluation follows each benchmark's native scoring convention: AIME24/25 averages the 2024 and 2025 scores, CFBench reports CSR/ISR/PSR, and the remaining benchmarks retain their native scales. For intervention-based methods, the activation rate is defined as the percentage of benchmark examples that open at least one sGDS episode.

As shown in Table~\ref{tab:singprobemed-capability}, \guard-Med activates on only 4.69\% of HealthBench examples and on none of the reported examples from AIME24/25, CFBench, Arena-Hard-v2, AlignBench, or WritingBench. Its capability scores remain broadly comparable to native AntAngelMed-100B across these evaluations. In contrast, always-on sGDS modifies every example and records lower scores than \guard-Med on all six benchmarks, with particularly noticeable degradation on AIME24/25 and Arena-Hard-v2.

\begin{table}[t]
    \centering
    \scriptsize
    \caption{Capability scores and sGDS activation rates (in parentheses).}
    \label{tab:singprobemed-capability}
    \resizebox{0.98\textwidth}{!}{%
    \begin{tabular}{@{}lcccccc@{}}
        \toprule
        Method & AIME24/25 avg. & HealthBench & CFBench & Arena-Hard-v2 & AlignBench & WritingBench \\
        \midrule
        AntAngelMed-100B & 66.67 & 88.80 & 84.31/59.00/69.50 & 52.06 & 8.435 & 8.351 \\
        Always-on sGDS & 60.00 (100\%) & 88.41 (100\%) & 81.73/52.00/64.00 (100\%) & 46.19 (100\%) & 8.465 (100\%) & 8.138 (100\%) \\
        SingProbe-Med & 66.67 (0\%) & 90.10 (4.69\%) & 84.31/59.00/69.50 (0\%) & 52.06 (0\%) & 8.435 (0\%) & 8.351 (0\%) \\
        \bottomrule
    \end{tabular}}
\end{table}

\begin{table}[t]
    \centering
    \small
    \caption{Request-level mean latency across inference configurations. Entries marked with $\sim$ are component-based estimates rather than directly measured end-to-end means.}
    \label{tab:singprobemed-latency}
    \begin{tabular}{@{}lr@{}}
        \toprule
        Configuration & Mean (s) \\
        \midrule
        AntAngelMed-100B, single branch & 37.319 \\
        AntAngelMed-100B $+$ SingProbe signals & $\sim$37.379 \\
        Full resident dual branch & 60.934 \\
        64-token on-demand window & $\sim$39.162 \\
        \bottomrule
    \end{tabular}
    \vspace{-1em}
\end{table}


\subsubsection{Inference Efficiency}

Finally, we evaluate whether on-demand intervention reduces the inference cost of dual-branch decoding. We compare four AntAngelMed-100B inference configurations: native single-branch decoding, single-branch decoding with continuous \guard-Med monitoring, a fully resident dual-branch configuration, and a bounded 64-token intervention window. The single-branch and resident dual-branch entries are reference measurements. The target-plus-\guard-Med and 64-token entries are component-based estimates derived from the measured single-branch latency and the corresponding additional signal or active-window cost, and are therefore marked with $\sim$ in Table~\ref{tab:singprobemed-latency}.

Continuous \guard-Med scoring contributes an estimated 0.060 seconds beyond the 37.319-second single-branch reference, yielding approximately 37.379 seconds before intervention cost is considered. Maintaining the full dual branch throughout generation increases the request-level mean latency to 60.934 seconds, an additional 23.615 seconds over native decoding. In comparison, a 64-token intervention window introduces an estimated 1.843-second active-window cost, giving approximately 39.162 seconds when added to the single-branch reference. These results show that bounding sGDS to the admitted suffix converts persistent dual-branch computation into a limited intervention cost. Together with the correction results above, they demonstrate that most of the safety benefit of full intervention can be retained without continuously executing the second branch throughout generation.

\section{Conclusion}

In this work, we introduced \guard, an intrinsic guardrail that directly leverages hidden states produced by the base LLM to jointly monitor query intent, response safety, and hallucination risk throughout generation. By reusing representations already computed during decoding, \guard avoids redundant text encoding and enables fine-grained streaming detection with substantially lower deployment complexity than standalone guardrail models. We also introduced SingStreamBench to evaluate the temporal behavior of streaming safety detectors, particularly their ability to remain silent on benign prefixes and identify the onset of unsafe content. Across extensive offline and online evaluations, \guard achieves strong performance on safety and hallucination detection while maintaining very low false-positive rates.

More importantly, our results suggest that the value of intrinsic guardrails extends beyond passive classification. \guard scores contain information about the safety tendency of future continuations and can therefore serve as control signals for constrained decoding. \guard-Med further demonstrates how such internal-state signals can support selective, on-demand intervention in a high-stakes domain, modifying generation only when and where intervention is warranted. Overall, \guard provides a lightweight interface between the internal representations of an LLM and generation-time safety mechanisms, offering a path toward guardrails that jointly improve monitoring granularity, computational efficiency, and actionable runtime control.

\section*{Authors}

Shiwen Cui, Jianjie Jiang, Guoyi Li, Jinzhen Lin, Changhua Meng, Shuo Shao, Jiashui Wang, Weiqiang Wang, Weixi Wu, Zhuoer Xu, Zihan Yan, Shenglin Yin, Xinlei Ying, Zhe Zhao, Jing Zhou.

\textit{(All authors are listed alphabetically by last name.)}

\section*{Author Contributions}

Shuo Shao, Zhuoer Xu, and Zihan Yan contributed to the safety aspect, and Jianjie Jiang, Weixi Wu, Shenglin Yin, Xinlei Ying, and Zhe Zhao contributed to the hallucination aspect of \guard. Jinzhen Lin developed the supporting infrastructure. Guoyi Li and Jing Zhou developed \guard-med. Shiwen Cui, Changhua Meng, Jiashui Wang, and Weiqiang Wang supervised this project.

\clearpage

\bibliographystyle{antgroup}
\bibliography{guard_ref}

@article{team2026singguard,
  title={SingGuard: A Policy-Adaptive Multimodal LLM Guardrail with Dynamic Reasoning},
  author={Ant Group},
  journal={arXiv preprint arXiv:2606.22873},
  year={2026}
}

@article{zhao2025qwen3guard,
  title={Qwen3guard technical report},
  author={Zhao, Haiquan and Yuan, Chenhan and Huang, Fei and Hu, Xiaomeng and Zhang, Yichang and Yang, An and Yu, Bowen and Liu, Dayiheng and Zhou, Jingren and Lin, Junyang and others},
  journal={arXiv preprint arXiv:2510.14276},
  year={2025}
}

@article{inan2023llamaguard,
  title={Llama guard: Llm-based input-output safeguard for human-ai conversations},
  author={Inan, Hakan and Upasani, Kartikeya and Chi, Jianfeng and Rungta, Rashi and Iyer, Krithika and Mao, Yuning and Tontchev, Michael and Hu, Qing and Fuller, Brian and Testuggine, Davide and others},
  journal={arXiv preprint arXiv:2312.06674},
  year={2023}
}

@article{lin2026yufengxguard,
  title={YuFeng-XGuard: A Reasoning-Centric, Interpretable, and Flexible Guardrail Model for Large Language Models},
  author={Lin, Junyu and Liu, Meizhen and Huang, Xiufeng and Li, Jinfeng and Hong, Haiwen and Yuan, Xiaohan and Chen, Yuefeng and Huang, Longtao and Xue, Hui and Duan, Ranjie and others},
  journal={arXiv preprint arXiv:2601.15588},
  year={2026}
}

@article{cunningham2026CC++,
  title={Constitutional Classifiers++: Efficient Production-Grade Defenses against Universal Jailbreaks},
  author={Cunningham, Hoagy and Wei, Jerry and Wang, Zihan and Persic, Andrew and Peng, Alwin and Abderrachid, Jordan and Agarwal, Raj and Chen, Bobby and Cohen, Austin and Dau, Andy and others},
  journal={arXiv preprint arXiv:2601.04603},
  year={2026}
}

@article{han2024wildguard,
  title={Wildguard: Open one-stop moderation tools for safety risks, jailbreaks, and refusals of llms},
  author={Han, Seungju and Rao, Kavel and Ettinger, Allyson and Jiang, Liwei and Lin, Bill Yuchen and Lambert, Nathan and Choi, Yejin and Dziri, Nouha},
  journal={Advances in neural information processing systems},
  volume={37},
  pages={8093--8131},
  year={2024}
}

@inproceedings{zheng2024sglang,
  author       = {Lianmin Zheng and
                  Liangsheng Yin and
                  Zhiqiang Xie and
                  Chuyue Sun and
                  Jeff Huang and
                  Cody Hao Yu and
                  Shiyi Cao and
                  Christos Kozyrakis and
                  Ion Stoica and
                  Joseph E. Gonzalez and
                  Clark W. Barrett and
                  Ying Sheng},
  title        = {SGLang: Efficient Execution of Structured Language Model Programs},
  booktitle    = {Advances in Neural Information Processing Systems 37},
  year         = {2024},
}

@misc{google2025gemini3,
  author        = {{Google}},
  title        = {Gemini 3},
  howpublished = {\url{https://blog.google/products-and-platforms/products/gemini/gemini-3/}},
  year         = {2025}
}

@misc{openai2025gpt5,
  author        = {{OpenAI}},
  title        = {GPT-5},
  howpublished = {\url{https://openai.com/index/gpt-5-1/}},
  year         = {2025}
}

@article{graniteguardian2024,
  title={Granite Guardian},
  author={IBM Research},
  journal={Technical Report},
  year={2024}
}

@article{zeng2024shieldgemma,
  title={ShieldGemma: Generative AI Content Moderation Based on Gemma},
  author={Zeng, Wenjun and others},
  journal={arXiv preprint arXiv:2407.21772},
  year={2024}
}

@article{li2025fineharm,
  title={From judgment to interference: Early stopping llm harmful outputs via streaming content monitoring},
  author={Li, Yang and Sheng, Qiang and Yang, Yehan and Zhang, Xueyao and Cao, Juan},
  journal={Advances in Neural Information Processing Systems},
  volume={38},
  pages={54305--54333},
  year={2025}
}

@article{ji2023beavertails,
  title={Beavertails: Towards improved safety alignment of llm via a human-preference dataset},
  author={Ji, Jiaming and Liu, Mickel and Dai, Josef and Pan, Xuehai and Zhang, Chi and Bian, Ce and Chen, Boyuan and Sun, Ruiyang and Wang, Yizhou and Yang, Yaodong},
  journal={Advances in Neural Information Processing Systems},
  volume={36},
  pages={24678--24704},
  year={2023}
}

@inproceedings{mazeika2024harmbench,
  title={Harmbench: A standardized evaluation framework for automated red teaming and robust refusal},
  author={Mazeika, Mantas and Phan, Long and Yin, Xuwang and Zou, Andy and Wang, Zifan and Mu, Norman and Sakhaee, Elham and Li, Nathaniel and Basart, Steven and Li, Bo and others},
  booktitle={International Conference on Machine Learning},
  year={2024}
}

@inproceedings{ghosh2025aegis2,
    title = "{AEGIS}2.0: A Diverse {AI} Safety Dataset and Risks Taxonomy for Alignment of {LLM} Guardrails",
    author = "Ghosh, Shaona and Varshney, Prasoon and Sreedhar, Makesh Narsimhan and Padmakumar, Aishwarya and Rebedea, Traian and Varghese, Jibin Rajan and Parisien, Christopher",
    editor = "Chiruzzo, Luis and Ritter, Alan and Wang, Lu",
    booktitle = "Proceedings of the 2025 Conference of the Nations of the Americas Chapter of the Association for Computational Linguistics: Human Language Technologies",
    month = apr,
    year = "2025",
    address = "Albuquerque, New Mexico",
    publisher = "Association for Computational Linguistics",
    url = "https://aclanthology.org/2025.naacl-long.306/",
    doi = "10.18653/v1/2025.naacl-long.306",
    pages = "5992--6026",
    ISBN = "979-8-89176-189-6"
}

@article{yang2025qwen3,
  title={Qwen3 technical report},
  author={Yang, An and Li, Anfeng and Yang, Baosong and Zhang, Beichen and Hui, Binyuan and Zheng, Bo and Yu, Bowen and Gao, Chang and Huang, Chengen and Lv, Chenxu and others},
  journal={arXiv preprint arXiv:2505.09388},
  year={2025}
}

@article{zeng2026glm5,
  title={Glm-5: from vibe coding to agentic engineering},
  author={Zeng, Aohan and Lv, Xin and Hou, Zhenyu and Du, Zhengxiao and Zheng, Qinkai and Chen, Bin and Yin, Da and Ge, Chendi and Huang, Chenghua and Xie, Chengxing and others},
  journal={arXiv preprint arXiv:2602.15763},
  year={2026}
}

@misc{kimik26,
  author       = {{Moonshot AI}},
  title        = {Kimi K2.6},
  year         = {2026},
  howpublished = {\url{https://huggingface.co/moonshotai/Kimi-K2.6}},
  note         = {Hugging Face model repository}
}

@misc{qwen3.5,
    title  = {{Qwen3.5}: Towards Native Multimodal Agents},
    author = {{Qwen Team}},
    month  = {February},
    year   = {2026},
    url    = {https://qwen.ai/blog?id=qwen3.5}
}

@inproceedings{markov2023openaimoderation,
  title={A holistic approach to undesired content detection in the real world},
  author={Markov, Todor and Zhang, Chong and Agarwal, Sandhini and Nekoul, Florentine Eloundou and Lee, Theodore and Adler, Steven and Jiang, Angela and Weng, Lilian},
  booktitle={Proceedings of the AAAI conference on artificial intelligence},
  volume={37},
  number={12},
  pages={15009--15018},
  year={2023}
}

@article{rottger2023xstest,
  title={Xstest: A test suite for identifying exaggerated safety behaviours in large language models},
  author={R{\"o}ttger, Paul and Kirk, Hannah Rose and Vidgen, Bertie and Attanasio, Giuseppe and Bianchi, Federico and Hovy, Dirk},
  journal={arXiv preprint arXiv:2308.01263},
  year={2023}
}

@inproceedings{
  choi2026expguard,
  title={ExpGuard: {LLM} Content Moderation in Specialized Domains},
  author={Choi, Minseok and Kim, Dongjin and Yang, Seungbin and Kim, Subin and Kwak, Youngjun and Oh, Juyoung and Choo, Jaegul and Son, Jungmin},
  booktitle={International Conference on Learning Representations},
  year={2026}
}

@article{ji2024pku,
  title={PKU-SafeRLHF: Towards Multi-Level Safety Alignment for LLMs with Human Preference},
  author={Ji, Jiaming and Hong, Donghai and Zhang, Borong and Chen, Boyuan and Dai, Josef and Zheng, Boren and Qiu, Tianyi and Li, Boxun and Yang, Yaodong},
  journal={arXiv preprint arXiv:2406.15513},
  year={2024}
}

@article{hendrycks2021math,
  title={Measuring Mathematical Problem Solving With the MATH Dataset},
  author={Dan Hendrycks and Collin Burns and Saurav Kadavath and Akul Arora and Steven Basart and Eric Tang and Dawn Song and Jacob Steinhardt},
  journal={NeurIPS},
  year={2021}
}

@misc{DatabricksBlog2023DollyV2,
    author    = {Mike Conover and Matt Hayes and Ankit Mathur and Jianwei Xie and Jun Wan and Sam Shah and Ali Ghodsi and Patrick Wendell and Matei Zaharia and Reynold Xin},
    title     = {Free Dolly: Introducing the World's First Truly Open Instruction-Tuned LLM},
    year      = {2023},
    url       = {https://www.databricks.com/blog/2023/04/12/dolly-first-open-commercially-viable-instruction-tuned-llm},
    urldate   = {2023-06-30}
}

@article{cobbe2021gsm8k,
  title={Training Verifiers to Solve Math Word Problems},
  author={Cobbe, Karl and Kosaraju, Vineet and Bavarian, Mohammad and Chen, Mark and Jun, Heewoo and Kaiser, Lukasz and Plappert, Matthias and Tworek, Jerry and Hilton, Jacob and Nakano, Reiichiro and Hesse, Christopher and Schulman, John},
  journal={arXiv preprint arXiv:2110.14168},
  year={2021}
}

@article{austin2021mbpp,
  title={Program Synthesis with Large Language Models},
  author={Austin, Jacob and Odena, Augustus and Nye, Maxwell and Bosma, Maarten and Michalewski, Henryk and Dohan, David and Jiang, Ellen and Cai, Carrie and Terry, Michael and Le, Quoc and others},
  journal={arXiv preprint arXiv:2108.07732},
  year={2021}
}

@inproceedings{OpenBookQA2018,
 title={Can a Suit of Armor Conduct Electricity? A New Dataset for Open Book Question Answering},
 author={Todor Mihaylov and Peter Clark and Tushar Khot and Ashish Sabharwal},
 booktitle={EMNLP},
 year={2018}
}

@article{chen2023factchd,
  title={Factchd: Benchmarking fact-conflicting hallucination detection},
  author={Chen, Xiang and Song, Duanzheng and Gui, Honghao and Wang, Chenxi and Zhang, Ningyu and Jiang, Yong and Huang, Fei and Lv, Chengfei and Zhang, Dan and Chen, Huajun},
  journal={arXiv preprint arXiv:2310.12086},
  year={2023}
}

@article{dziri2022faithdial,
  title={Faithdial: A faithful benchmark for information-seeking dialogue},
  author={Dziri, Nouha and Kamalloo, Ehsan and Milton, Sivan and Za{\"\i}ane, Osmar R and Yu, Mo and Ponti, Edoardo M and Reddy, Siva},
  journal={Transactions of the Association for Computational Linguistics},
  volume={10},
  pages={1473--1490},
  year={2022}
}

@article{mishra2024fava,
  title={Fine-grained hallucination detection and editing for language models},
  author={Mishra, Abhika and Asai, Akari and Balachandran, Vidhisha and Wang, Yizhong and Neubig, Graham and Tsvetkov, Yulia and Hajishirzi, Hannaneh},
  journal={arXiv preprint arXiv:2401.06855},
  year={2024}
}

@inproceedings{niu2024ragtruth,
  title={Ragtruth: A hallucination corpus for developing trustworthy retrieval-augmented language models},
  author={Niu, Cheng and Wu, Yuanhao and Zhu, Juno and Xu, Siliang and Shum, Kashun and Zhong, Randy and Song, Juntong and Zhang, Tong},
  booktitle={Proceedings of the 62nd Annual Meeting of the Association for Computational Linguistics},
  pages={10862--10878},
  year={2024}
}

@inproceedings{mickus2024semeval,
  title={SemEval-2024 task 6: SHROOM, a shared-task on hallucinations and related observable overgeneration mistakes},
  author={Mickus, Timothee and Zosa, Elaine and V{\'a}zquez, Ra{\'u}l and Vahtola, Teemu and Tiedemann, J{\"o}rg and Segonne, Vincent and Raganato, Alessandro and Apidianaki, Marianna},
  booktitle={Proceedings of the 18th International Workshop on Semantic Evaluation},
  pages={1979--1993},
  year={2024}
}

@inproceedings{manakul2023selfcheckgpt,
  title={Selfcheckgpt: Zero-resource black-box hallucination detection for generative large language models},
  author={Manakul, Potsawee and Liusie, Adian and Gales, Mark},
  booktitle={Proceedings of the 2023 conference on empirical methods in natural language processing},
  pages={9004--9017},
  year={2023}
}

@article{agarwal2025gptoss,
  title={gpt-oss-120b \& gpt-oss-20b model card},
  author={Agarwal, Sandhini and Ahmad, Lama and Ai, Jason and Altman, Sam and Applebaum, Andy and Arbus, Edwin and Arora, Rahul K and Bai, Yu and Baker, Bowen and Bao, Haiming and others},
  journal={arXiv preprint arXiv:2508.10925},
  year={2025}
}

@article{bhatnagar2026drift,
  title={DRIFT: Detecting Representational Inconsistencies for Factual Truthfulness},
  author={Bhatnagar, Rohan and Sun, Youran and Zhang, Chi Andrew and Wen, Yixin and Yang, Haizhao},
  journal={arXiv preprint arXiv:2601.14210},
  year={2026}
}

@inproceedings{azaria2023internal,
  title={The internal state of an LLM knows when it’s lying},
  author={Azaria, Amos and Mitchell, Tom},
  booktitle={Findings of the Association for Computational Linguistics: EMNLP 2023},
  pages={967--976},
  year={2023}
}

@article{niu2025robust,
  title={Robust hallucination detection in llms via adaptive token selection},
  author={Niu, Mengjia and Haddadi, Hamed and Pang, Guansong},
  journal={Advances in Neural Information Processing Systems},
  volume={38},
  pages={126355--126377},
  year={2025}
}

@article{jackson2025aaomniscience,
  title={AA-Omniscience: Evaluating cross-domain knowledge reliability in large language models},
  author={Jackson, Declan and Keating, William and Cameron, George and Hill-Smith, Micah},
  journal={arXiv preprint arXiv:2511.13029},
  year={2025}
}

@inproceedings{suzgun2023bbh,
  title={Challenging big-bench tasks and whether chain-of-thought can solve them},
  author={Suzgun, Mirac and Scales, Nathan and Sch{\"a}rli, Nathanael and Gehrmann, Sebastian and Tay, Yi and Chung, Hyung Won and Chowdhery, Aakanksha and Le, Quoc and Chi, Ed H and Zhou, Denny and others},
  booktitle={Findings of the Association for Computational Linguistics: ACL 2023},
  pages={13003--13051},
  year={2023}
}

@inproceedings{lee2019nqopen,
  title={Latent retrieval for weakly supervised open domain question answering},
  author={Lee, Kenton and Chang, Ming-Wei and Toutanova, Kristina},
  booktitle={Proceedings of the 57th annual meeting of the association for computational linguistics},
  pages={6086--6096},
  year={2019}
}

@inproceedings{rajpurkar2016squad,
  title={Squad: 100,000+ questions for machine comprehension of text},
  author={Rajpurkar, Pranav and Zhang, Jian and Lopyrev, Konstantin and Liang, Percy},
  booktitle={Proceedings of the 2016 conference on empirical methods in natural language processing},
  pages={2383--2392},
  year={2016}
}

@inproceedings{lin2022truthfulqa,
  title={Truthfulqa: Measuring how models mimic human falsehoods},
  author={Lin, Stephanie and Hilton, Jacob and Evans, Owain},
  booktitle={Proceedings of the 60th annual meeting of the association for computational linguistics (volume 1: long papers)},
  pages={3214--3252},
  year={2022}
}

@article{singhal2023large,
  title = {Large Language Models Encode Clinical Knowledge},
  author = {Singhal, Karan and Azizi, Shekoofeh and Tu, Tao and Mahdavi, S. Sara and Wei, Jason and Chung, Hyung Won and Scales, Nathan and Tanwani, Ajay and Cole-Lewis, Heather and Pfohl, Stephen and others},
  journal = {Nature},
  volume = {620},
  number = {7972},
  pages = {172--180},
  year = {2023},
  doi = {10.1038/s41586-023-06291-2}
}

@article{mishra2024finegrained,
  title = {Fine-Grained Hallucination Detection and Editing for Language Models},
  author = {Mishra, Abhika and Asai, Akari and Balachandran, Vidhisha and Wang, Yizhong and Neubig, Graham and Tsvetkov, Yulia and Hajishirzi, Hannaneh},
  journal = {arXiv preprint arXiv:2401.06855},
  year = {2024}
}

@article{yang2024medguard,
  title = {Ensuring Safety and Trust: Analyzing the Risks of Large Language Models in Medicine},
  author = {Yang, Yifan and Jin, Qiao and Leaman, Robert and Liu, Xiaoyu and Xiong, Guangzhi and Sarfo-Gyamfi, Maame and Gong, Changlin and Ferri{\`e}re-Steinert, Santiago and Wilbur, W. John and Li, Xiaojun and Yuan, Jiaxin and An, Bang and Castro, Kelvin S. and others},
  journal = {arXiv preprint arXiv:2411.14487},
  year = {2024},
  doi = {10.48550/arXiv.2411.14487}
}

@article{draelos2026unsafe,
  title = {Large Language Models Provide Unsafe Answers to Patient-Posed Medical Questions},
  author = {Draelos, Rachel L. and Afreen, Samina and Blasko, Barbara and Brazile, Tiffany L. and Chase, Natasha and Desai, Dimple Patel and Evert, Jessica and Gardner, Heather L. and Herrmann, Lauren and House, Aswathy Vaikom and Kass, Stephanie and Kavan, Marianne and Khemani, Kirshma and Koire, Amanda and McDonald, Lauren M. and Rabeeah, Zahraa and Shah, Amy},
  journal = {npj Digital Medicine},
  volume = {9},
  number = {1},
  pages = {241},
  year = {2026},
  doi = {10.1038/s41746-026-02428-5},
  url = {https://www.nature.com/articles/s41746-026-02428-5}
}

@misc{AntAngelMed,
  title  = {{AntAngelMed}: A High-Performance Medical Language Model with Efficient {MoE}-Powered Clinical Reasoning},
  author = {{AntAngelMed Team}},
  year   = {2025},
  url    = {https://huggingface.co/MedAIBase/AntAngelMed}
}

@misc{maxwelljia2024aime,
  title        = {AIME 2024 Dataset},
  author       = {{Maxwell-Jia}},
  year         = {2024},
  howpublished = {Hugging Face Dataset},
  url          = {https://huggingface.co/datasets/Maxwell-Jia/AIME_2024}
}

@misc{aime25,
  title     = {American Invitational Mathematics Examination (AIME) 2025},
  author    = {Zhang, Yifan and {Math-AI Team}},
  year      = {2025},
  publisher = {Hugging Face},
  url       = {https://huggingface.co/datasets/math-ai/aime25}
}

@inproceedings{zhang2025cfbench,
  title={Cfbench: A comprehensive constraints-following benchmark for llms},
  author={Zhang, Tao and Zhu, Chenglin and Shen, Yanjun and Luo, Wenjing and Zhang, Yan and Liang, Hao and Yang, Fan and Lin, Mingan and Qiao, Yujing and Chen, Weipeng and others},
  booktitle={Annual Meeting of the Association for Computational Linguistics},
  pages={32926--32944},
  year={2025}
}

@article{arora2025healthbench,
  title={Healthbench: Evaluating large language models towards improved human health},
  author={Arora, Rahul K and Wei, Jason and Hicks, Rebecca Soskin and Bowman, Preston and Qui{\~n}onero-Candela, Joaquin and Tsimpourlas, Foivos and Sharman, Michael and Shah, Meghan and Vallone, Andrea and Beutel, Alex and others},
  journal={arXiv preprint arXiv:2505.08775},
  year={2025}
}

@inproceedings{li2025arenahard,
  title     = {From Crowdsourced Data to High-quality Benchmarks:
               Arena-Hard and BenchBuilder Pipeline},
  author    = {Li, Tianle and Chiang, Wei-Lin and Frick, Evan and Dunlap, Lisa and
               Wu, Tianhao and Zhu, Banghua and Gonzalez, Joseph E. and Stoica, Ion},
  booktitle = {Proceedings of the 42nd International Conference on Machine Learning},
  series    = {Proceedings of Machine Learning Research},
  volume    = {267},
  pages     = {34209--34231},
  publisher = {PMLR},
  year      = {2025}
}

@inproceedings{liu2024alignbench,
  title={Alignbench: Benchmarking chinese alignment of large language models},
  author={Liu, Xiao and Lei, Xuanyu and Wang, Shengyuan and Huang, Yue and Feng, Andrew and Wen, Bosi and Cheng, Jiale and Ke, Pei and Xu, Yifan and Tam, Weng Lam and others},
  booktitle={Annual Meeting of the Association for Computational Linguistics},
  pages={11621--11640},
  year={2024}
}

@article{wu2026writingbench,
  title={Writingbench: A comprehensive benchmark for generative writing},
  author={Wu, Yuning and Mei, Jiahao and Yan, Ming and Li, Chenliang and Lai, Shaopeng and Ren, Yuran and Wang, Zijia and Zhang, Ji and Wu, Mengyue and Jin, Qin and others},
  journal={Advances in Neural Information Processing Systems},
  volume={38},
  year={2026}
}

@inproceedings{hu2022lora,
  title={LoRA: Low-Rank Adaptation of Large Language Models},
  author={Hu, Edward J. and Shen, Yelong and Wallis, Phillip and Allen-Zhu, Zeyuan and Li, Yuanzhi and Wang, Shean and Wang, Lu and Chen, Weizhu},
  booktitle={International Conference on Learning Representations},
  year={2022}
}

@inproceedings{kwon2023vllm,
  title={Efficient memory management for large language model serving with pagedattention},
  author={Kwon, Woosuk and Li, Zhuohan and Zhuang, Siyuan and Sheng, Ying and Zheng, Lianmin and Yu, Cody Hao and Gonzalez, Joseph and Zhang, Hao and Stoica, Ion},
  booktitle={Proceedings of the 29th symposium on operating systems principles},
  pages={611--626},
  year={2023}
}

@misc{qwen36_35b_a3b,
    title = {{Qwen3.6-35B-A3B}: Agentic Coding Power, Now Open to All},
    url = {https://qwen.ai/blog?id=qwen3.6-35b-a3b},
    author = {{Qwen Team}},
    month = {April},
    year = {2026}
}

@misc{qwen3.6-27b,
    title  = {{Qwen3.6-27B}: Flagship-Level Coding in a {27B} Dense Model},
    author = {{Qwen Team}},
    month  = {April},
    year   = {2026},
    url    = {https://qwen.ai/blog?id=qwen3.6-27b}
}

@article{grattafiori2024llama3,
  title={The llama 3 herd of models},
  author={Grattafiori, Aaron and Dubey, Abhimanyu and Jauhri, Abhinav and Pandey, Abhinav and Kadian, Abhishek and Al-Dahle, Ahmad and Letman, Aiesha and Mathur, Akhil and Schelten, Alan and Vaughan, Alex and others},
  journal={arXiv preprint arXiv:2407.21783},
  year={2024}
}

@article{team2026gemma4,
  title={Gemma 4 technical report},
  author={Team, Gemma and Abd, Sherif El and Aggarwal, Vaibhav and Algayres, Robin and Andreev, Alek and Bachem, Olivier and Ballantyne, Ian and Brick, Cormac and C{\u{a}}rbune, Victor and Casbon, Michelle and others},
  journal={arXiv preprint arXiv:2607.02770},
  year={2026}
}

@article{xu2026deepseekv4,
  title={Deepseek-v4: Towards highly efficient million-token context intelligence},
  author={Xu, Anyi and Lin, Bangcai and Xue, Bing and Wang, Bingxuan and Xu, Bingzheng and Wu, Bochao and Zhang, Bowei and Lin, Chaofan and Dong, Chen and Ling, Chenchen and others},
  journal={arXiv preprint arXiv:2606.19348},
  year={2026}
}

@misc{stepfun2026step37flash,
  author       = {{StepFun Team}},
  title        = {Step-3.7-Flash},
  year         = {2026},
  howpublished = {Hugging Face model repository},
  url          = {https://huggingface.co/stepfun-ai/Step-3.7-Flash},
  note         = {Accessed September 8, 2026}
}

@article{chen2026minimaxm2,
  title         = {The {MiniMax-M2} Series: Mini Activations Unleashing Max Real-World Intelligence},
  author        = {Chen, Aili and others},
  journal       = {arXiv preprint arXiv:2605.26494},
  year          = {2026},
  eprint        = {2605.26494},
  archivePrefix = {arXiv},
  primaryClass  = {cs.AI},
  doi           = {10.48550/arXiv.2605.26494},
  url           = {https://arxiv.org/abs/2605.26494}
}

@misc{tencent2026hy3,
  author       = {{Tencent Hy Team}},
  title        = {{Hy3}},
  year         = {2026},
  howpublished = {Hugging Face model repository},
  url          = {https://huggingface.co/tencent/Hy3},
  note         = {Accessed September 8, 2026}
}

@article{li2025kelp,
  title={Kelp: A streaming safeguard for large models via latent dynamics-guided risk detection},
  author={Li, Xiaodan and Wu, Mengjie and Zhu, Yao and Lv, Yunna and Chen, YueFeng and Chen, Cen and Guo, Jianmei and Xue, Hui},
  journal={arXiv preprint arXiv:2510.09694},
  year={2025}
}

\clearpage

\appendix
\section*{Appendix}

\section{Adaptation to More Open-Source Base Models}
\label{app:adaptation}
We extend the availability of \guard by adapting SingProbe to a broad set of mainstream open-source base models. The adapted models include the Qwen3~\citep{yang2025qwen3}, Qwen3.5~\citep{qwen3.5}, Qwen3.6~\citep{qwen3.6-27b, qwen36_35b_a3b}, and Qwen3.8 series; the Llama 3~\citep{grattafiori2024llama3} series; the Gemma 4~\citep{team2026gemma4} series; the Ling 3.0 series; the gpt-oss~\citep{agarwal2025gptoss} series; the GLM~\citep{zeng2026glm5} series; and representative models from the StepFun~\citep{stepfun2026step37flash}, MiniMax~\citep{chen2026minimaxm2}, DeepSeek~\citep{xu2026deepseekv4}, and Hy~\citep{tencent2026hy3} families. This effort makes the same SingProbe interface available across commonly used model families, parameter scales, and architectures.

Table~\ref{tab:cross_model_adaptation} lists the adapted base models and their corresponding SingProbe repositories. Each probe is intended for use with the paired base-model repository shown in the table.

\begin{table}[h]
\centering
\scriptsize
\caption{SingProbe adaptation and release coverage across open-source base models. Base Model and Probe Model report Hugging Face repository IDs.}
\vspace{0.5mm}
\label{tab:cross_model_adaptation}
\resizebox{0.95\textwidth}{!}{%
\renewcommand{\arraystretch}{1.10}
\setlength{\tabcolsep}{4.2pt}
\begin{tabular}{llcc}
\toprule[1.2pt]
\textbf{Base Model} & \textbf{Probe Model} & \textbf{Base params} & \textbf{Probe params} \\
\midrule[0.8pt]
\href{https://huggingface.co/Qwen/Qwen3-0.6B}{\nolinkurl{Qwen/Qwen3-0.6B}} & \href{https://huggingface.co/inclusionAI/Qwen3-0.6B-singprobe}{\nolinkurl{inclusionAI/Qwen3-0.6B-singprobe}} & 0.6B & 2.23M \\
\href{https://huggingface.co/Qwen/Qwen3.5-0.8B}{\nolinkurl{Qwen/Qwen3.5-0.8B}} & \href{https://huggingface.co/inclusionAI/Qwen3.5-0.8B-singprobe}{\nolinkurl{inclusionAI/Qwen3.5-0.8B-singprobe}} & 0.8B & 2.23M \\
\href{https://huggingface.co/meta-llama/Llama-3.2-1B-Instruct}{\nolinkurl{meta-llama/Llama-3.2-1B-Instruct}} & \href{https://huggingface.co/inclusionAI/Llama-3.2-1B-Instruct-singprobe}{\nolinkurl{inclusionAI/Llama-3.2-1B-Instruct-singprobe}} & 1.3B & 4.2M \\
\href{https://huggingface.co/Qwen/Qwen3.5-2B}{\nolinkurl{Qwen/Qwen3.5-2B}} & \href{https://huggingface.co/inclusionAI/Qwen3.5-2B-singprobe}{\nolinkurl{inclusionAI/Qwen3.5-2B-singprobe}} & 2.0B & 4.2M \\
\href{https://huggingface.co/google/gemma-4-E4B-it}{\nolinkurl{google/gemma-4-E4B-it}} & \href{https://huggingface.co/inclusionAI/gemma-4-E4B-it-singprobe}{\nolinkurl{inclusionAI/gemma-4-E4B-it-singprobe}} & 4.0B & 5.18M \\
\href{https://huggingface.co/Qwen/Qwen3-4B-Instruct-2507}{\nolinkurl{Qwen/Qwen3-4B-Instruct-2507}} & \href{https://huggingface.co/inclusionAI/Qwen3-4B-Instruct-2507-singprobe}{\nolinkurl{inclusionAI/Qwen3-4B-Instruct-2507-singprobe}} & 4.0B & 5.18M \\
\href{https://huggingface.co/Qwen/Qwen3.5-4B}{\nolinkurl{Qwen/Qwen3.5-4B}} & \href{https://huggingface.co/inclusionAI/Qwen3.5-4B-singprobe}{\nolinkurl{inclusionAI/Qwen3.5-4B-singprobe}} & 4.0B & 5.18M \\
\href{https://huggingface.co/inclusionAI/Ling-3.0-tiny}{\nolinkurl{inclusionAI/Ling-3.0-tiny}} & \href{https://huggingface.co/inclusionAI/Ling-3.0-tiny-singprobe}{\nolinkurl{inclusionAI/Ling-3.0-tiny-singprobe}} & 7.9B & 3.22M \\
\href{https://huggingface.co/meta-llama/Llama-3.1-8B-Instruct}{\nolinkurl{meta-llama/Llama-3.1-8B-Instruct}} & \href{https://huggingface.co/inclusionAI/Llama-3.1-8B-Instruct-singprobe}{\nolinkurl{inclusionAI/Llama-3.1-8B-Instruct-singprobe}} & 8.0B & 8.13M \\
\href{https://huggingface.co/Qwen/Qwen3-8B}{\nolinkurl{Qwen/Qwen3-8B}} & \href{https://huggingface.co/inclusionAI/Qwen3-8B-singprobe}{\nolinkurl{inclusionAI/Qwen3-8B-singprobe}} & 8.0B & 8.13M \\
\href{https://huggingface.co/Qwen/Qwen3.5-9B}{\nolinkurl{Qwen/Qwen3.5-9B}} & \href{https://huggingface.co/inclusionAI/Qwen3.5-9B-singprobe}{\nolinkurl{inclusionAI/Qwen3.5-9B-singprobe}} & 9.0B & 8.13M \\
\href{https://huggingface.co/openai/gpt-oss-20b}{\nolinkurl{openai/gpt-oss-20b}} & \href{https://huggingface.co/inclusionAI/gpt-oss-20b-singprobe}{\nolinkurl{inclusionAI/gpt-oss-20b-singprobe}} & 20B & 5.8M \\
\href{https://huggingface.co/google/gemma-4-26B-A4B-it}{\nolinkurl{google/gemma-4-26B-A4B-it}} & \href{https://huggingface.co/inclusionAI/gemma-4-26B-A4B-it-singprobe}{\nolinkurl{inclusionAI/gemma-4-26B-A4B-it-singprobe}} & 26B & 5.67M \\
\href{https://huggingface.co/Qwen/Qwen3.5-27B}{\nolinkurl{Qwen/Qwen3.5-27B}} & \href{https://huggingface.co/inclusionAI/Qwen3.5-27B-singprobe}{\nolinkurl{inclusionAI/Qwen3.5-27B-singprobe}} & 27B & 10.1M \\
\href{https://huggingface.co/Qwen/Qwen3.6-27B}{\nolinkurl{Qwen/Qwen3.6-27B}} & \href{https://huggingface.co/inclusionAI/Qwen3.6-27B-singprobe}{\nolinkurl{inclusionAI/Qwen3.6-27B-singprobe}} & 27B & 10.1M \\
\href{https://huggingface.co/Qwen/Qwen3.8-27B}{\nolinkurl{Qwen/Qwen3.8-27B}} & \href{https://huggingface.co/inclusionAI/Qwen3.8-27B-singprobe}{\nolinkurl{inclusionAI/Qwen3.8-27B-singprobe}} & 27B & 10.1M \\
\href{https://huggingface.co/google/gemma-4-31B-it}{\nolinkurl{google/gemma-4-31B-it}} & \href{https://huggingface.co/inclusionAI/gemma-4-31B-it-singprobe}{\nolinkurl{inclusionAI/gemma-4-31B-it-singprobe}} & 31B & 10.59M \\
\href{https://huggingface.co/Qwen/Qwen3.5-35B-A3B}{\nolinkurl{Qwen/Qwen3.5-35B-A3B}} & \href{https://huggingface.co/inclusionAI/Qwen3.5-35B-A3B-singprobe}{\nolinkurl{inclusionAI/Qwen3.5-35B-A3B-singprobe}} & 35B & 4.2M \\
\href{https://huggingface.co/Qwen/Qwen3.6-35B-A3B}{\nolinkurl{Qwen/Qwen3.6-35B-A3B}} & \href{https://huggingface.co/inclusionAI/Qwen3.6-35B-A3B-singprobe}{\nolinkurl{inclusionAI/Qwen3.6-35B-A3B-singprobe}} & 35B & 4.2M \\
\href{https://huggingface.co/openai/gpt-oss-120b}{\nolinkurl{openai/gpt-oss-120b}} & \href{https://huggingface.co/inclusionAI/gpt-oss-120b-singprobe}{\nolinkurl{inclusionAI/gpt-oss-120b-singprobe}} & 120B & 5.8M \\
\href{https://huggingface.co/Qwen/Qwen3.5-122B-A10B}{\nolinkurl{Qwen/Qwen3.5-122B-A10B}} & \href{https://huggingface.co/inclusionAI/Qwen3.5-122B-A10B-singprobe}{\nolinkurl{inclusionAI/Qwen3.5-122B-A10B-singprobe}} & 122B & 6.17M \\
\href{https://huggingface.co/inclusionAI/Ling-3.0-flash}{\nolinkurl{inclusionAI/Ling-3.0-flash}} & \href{https://huggingface.co/inclusionAI/Ling-3.0-flash-singprobe}{\nolinkurl{inclusionAI/Ling-3.0-flash-singprobe}} & 124B & 5.18M \\
\href{https://huggingface.co/stepfun-ai/Step-3.7-Flash}{\nolinkurl{stepfun-ai/Step-3.7-Flash}} & \href{https://huggingface.co/inclusionAI/Step-3.7-Flash-singprobe}{\nolinkurl{inclusionAI/Step-3.7-Flash-singprobe}} & 196B & 8.13M \\
\href{https://huggingface.co/MiniMaxAI/MiniMax-M2.7}{\nolinkurl{MiniMaxAI/MiniMax-M2.7}} & \href{https://huggingface.co/inclusionAI/MiniMax-M2.7-singprobe}{\nolinkurl{inclusionAI/MiniMax-M2.7-singprobe}} & 229B & 6.17M \\
\href{https://huggingface.co/deepseek-ai/DeepSeek-V4-Flash-0731}{\nolinkurl{deepseek-ai/DeepSeek-V4-Flash-0731}} & \href{https://huggingface.co/inclusionAI/DeepSeek-V4-Flash-0731-singprobe}{\nolinkurl{inclusionAI/DeepSeek-V4-Flash-0731-singprobe}} & 284B & 8.13M \\
\href{https://huggingface.co/tencent/Hy3}{\nolinkurl{tencent/Hy3}} & \href{https://huggingface.co/inclusionAI/Hy3-singprobe}{\nolinkurl{inclusionAI/Hy3-singprobe}} & 295B & 8.13M \\
\href{https://huggingface.co/Qwen/Qwen3.5-397B-A17B}{\nolinkurl{Qwen/Qwen3.5-397B-A17B}} & \href{https://huggingface.co/inclusionAI/Qwen3.5-397B-A17B-singprobe}{\nolinkurl{inclusionAI/Qwen3.5-397B-A17B-singprobe}} & 397B & 8.13M \\
\href{https://huggingface.co/zai-org/GLM-5.2}{\nolinkurl{zai-org/GLM-5.2}} & \href{https://huggingface.co/inclusionAI/GLM-5.2-singprobe}{\nolinkurl{inclusionAI/GLM-5.2-singprobe}} & 753B & 12.06M \\
\href{https://huggingface.co/zai-org/GLM-5.3}{\nolinkurl{zai-org/GLM-5.3}} & \href{https://huggingface.co/inclusionAI/GLM-5.3-singprobe}{\nolinkurl{inclusionAI/GLM-5.3-singprobe}} & 753B & 12.06M \\
\bottomrule[1.2pt]
\end{tabular}
}
\end{table}

\section{Evaluation of Adapted SingProbe Models}
\label{app:cross_model_evaluation}
We evaluate the adapted probes with the same benchmark suites used in the main text. Table~\ref{tab:cross_model_summary} summarizes the headline metrics, followed by detailed results for query classification, response safety, and response hallucination.

Across the adapted releases, query F1 ranges from 0.8334 to 0.8811, response-safety F1 from 0.8334 to 0.8729, response-level streaming AUC from 0.9740 to 0.9905, and token-level streaming AUC from 0.9175 to 0.9508. The pooled benign false-positive rate remains at or below 0.08\% for 27 of the 29 models.

\begin{table}[t]
\centering
\scriptsize
\caption{Evaluation summary for the adapted SingProbe models. Query F1 and response-safety F1 are macro averages over six and eight benchmarks, respectively. Streaming metrics are averages over three benchmarks. Benign FPR is pooled over five benign datasets (lower is better), and hallucination AUC is the macro average over six benchmarks.}
\vspace{0.5mm}
\label{tab:cross_model_summary}
\resizebox{0.95\textwidth}{!}{%
\renewcommand{\arraystretch}{1.10}
\setlength{\tabcolsep}{3.8pt}
\begin{tabular}{lcccccc}
\toprule[1.2pt]
\textbf{Probe Model} & \textbf{Query F1} & \textbf{Resp. Safety F1} & \textbf{Stream R-AUC} & \textbf{Stream T-AUC} & \textbf{Benign FPR $\downarrow$} & \textbf{Hallu AUC} \\
\midrule[0.8pt]
Qwen3-0.6B-singprobe & 0.8400 & 0.8356 & 0.9819 & 0.9410 & 0.21\% & 0.7592 \\
Qwen3.5-0.8B-singprobe & 0.8334 & 0.8334 & 0.9778 & 0.9206 & 0.23\% & 0.7480 \\
Llama-3.2-1B-Instruct-singprobe & 0.8566 & 0.8468 & 0.9817 & 0.9242 & 0.08\% & 0.7546 \\
Qwen3.5-2B-singprobe & 0.8542 & 0.8516 & 0.9824 & 0.9243 & 0.08\% & 0.7642 \\
gemma-4-E4B-it-singprobe & 0.8443 & 0.8681 & 0.9870 & 0.9263 & 0.04\% & 0.7776 \\
Qwen3-4B-Instruct-2507-singprobe & 0.8640 & 0.8592 & 0.9879 & 0.9468 & 0.06\% & 0.7921 \\
Qwen3.5-4B-singprobe & 0.8650 & 0.8646 & 0.9870 & 0.9286 & 0.04\% & 0.7855 \\
Ling-3.0-tiny-singprobe & 0.8561 & 0.8508 & 0.9888 & 0.9479 & 0.07\% & 0.7765 \\
Llama-3.1-8B-Instruct-singprobe & 0.8759 & 0.8624 & 0.9858 & 0.9291 & 0.03\% & 0.7704 \\
Qwen3-8B-singprobe & 0.8687 & 0.8616 & 0.9878 & 0.9484 & 0.03\% & 0.7849 \\
Qwen3.5-9B-singprobe & 0.8719 & 0.8695 & 0.9874 & 0.9344 & 0.04\% & 0.7954 \\
gpt-oss-20b-singprobe & 0.8685 & 0.8644 & 0.9781 & 0.9338 & 0.05\% & 0.7581 \\
gemma-4-26B-A4B-it-singprobe & 0.8439 & 0.8647 & 0.9740 & 0.9175 & 0.04\% & 0.7952 \\
Qwen3.5-27B-singprobe & 0.8741 & \textbf{0.8729} & 0.9879 & 0.9289 & 0.03\% & 0.7967 \\
Qwen3.6-27B-singprobe & 0.8718 & 0.8692 & 0.9864 & 0.9270 & 0.02\% & 0.8094 \\
Qwen3.8-27B-singprobe & 0.8683 & 0.8676 & 0.9881 & 0.9305 & 0.03\% & 0.8035 \\
gemma-4-31B-it-singprobe & 0.8700 & 0.8655 & 0.9830 & 0.9184 & 0.03\% & 0.8054 \\
Qwen3.5-35B-A3B-singprobe & 0.8751 & 0.8672 & 0.9902 & 0.9374 & 0.03\% & 0.8117 \\
Qwen3.6-35B-A3B-singprobe & 0.8699 & 0.8658 & 0.9889 & 0.9359 & 0.03\% & 0.8023 \\
gpt-oss-120b-singprobe & 0.8699 & 0.8652 & 0.9881 & 0.9454 & 0.06\% & 0.7815 \\
Qwen3.5-122B-A10B-singprobe & 0.8697 & 0.8680 & 0.9895 & 0.9304 & 0.03\% & 0.7806 \\
Ling-3.0-flash-singprobe & 0.8674 & 0.8728 & 0.9887 & 0.9481 & 0.03\% & 0.8012 \\
Step-3.7-Flash-singprobe & 0.8502 & 0.8555 & 0.9858 & 0.9295 & 0.05\% & 0.7904 \\
MiniMax-M2.7-singprobe & 0.8731 & 0.8698 & 0.9902 & 0.9483 & 0.03\% & 0.7777 \\
DeepSeek-V4-Flash-0731-singprobe & \textbf{0.8811} & 0.8697 & 0.9882 & \textbf{0.9508} & \textbf{0.01\%} & 0.7949 \\
Hy3-singprobe & 0.8721 & 0.8645 & 0.9899 & 0.9498 & 0.02\% & 0.8067 \\
Qwen3.5-397B-A17B-singprobe & 0.8750 & 0.8696 & \textbf{0.9905} & 0.9339 & 0.03\% & 0.8117 \\
GLM-5.2-singprobe & 0.8677 & 0.8695 & 0.9852 & 0.9308 & 0.02\% & 0.8118 \\
GLM-5.3-singprobe & 0.8777 & 0.8715 & 0.9877 & 0.9322 & 0.03\% & \textbf{0.8170} \\
\bottomrule[1.2pt]
\end{tabular}
}
\end{table}

\subsection{Query Intent Classification}
Table~\ref{tab:cross_model_query} reports query-intent F1 for every benchmark. DeepSeek-V4-Flash-0731-singprobe achieves the highest macro F1 (0.8811), while Llama-3.1-8B-Instruct-singprobe, Qwen3.5-35B-A3B-singprobe, and Qwen3.5-397B-A17B-singprobe also exceed 0.8750. Results do not increase monotonically with parameter count: several compact models remain competitive, including Qwen3-4B-Instruct-2507-singprobe (0.8640) and Qwen3.5-4B-singprobe (0.8650). The OpenAI Moderation split is generally the most challenging, whereas XSTest and HarmBench yield consistently high F1.

\begin{table}[t]
\centering
\scriptsize
\caption{Query-intent classification F1 by benchmark. Macro is the unweighted mean over the six datasets.}
\vspace{0.5mm}
\label{tab:cross_model_query}
\resizebox{0.95\textwidth}{!}{%
\renewcommand{\arraystretch}{1.10}
\setlength{\tabcolsep}{3.8pt}
\begin{tabular}{lcccccc>{\columncolor{green!8}}c}
\toprule[1.2pt]
\textbf{Model} & \textbf{Aegis2} & \textbf{XGuard Test} & \textbf{OpenAI Moderation} & \textbf{XSTest} & \textbf{HarmBench} & \textbf{ExpGuardTest} & \textbf{Avg.} \\
\midrule[0.8pt]
Qwen3-0.6B-singprobe & 0.8409 & 0.8594 & 0.6810 & 0.8796 & 0.9094 & 0.8695 & 0.8400 \\
Qwen3.5-0.8B-singprobe & 0.8398 & 0.8654 & 0.6869 & 0.8668 & 0.8899 & 0.8515 & 0.8334 \\
Llama-3.2-1B-Instruct-singprobe & 0.8496 & 0.8730 & 0.7180 & 0.9032 & 0.9247 & 0.8708 & 0.8566 \\
Qwen3.5-2B-singprobe & 0.8450 & 0.8672 & 0.7339 & 0.9146 & 0.9089 & 0.8558 & 0.8542 \\
gemma-4-E4B-it-singprobe & 0.8266 & 0.8548 & 0.7439 & 0.8404 & 0.9211 & 0.8791 & 0.8443 \\
Qwen3-4B-Instruct-2507-singprobe & 0.8468 & 0.8782 & 0.7460 & 0.9051 & 0.9279 & 0.8800 & 0.8640 \\
Qwen3.5-4B-singprobe & 0.8550 & 0.8787 & 0.7348 & 0.9246 & 0.9288 & 0.8678 & 0.8650 \\
Ling-3.0-tiny-singprobe & 0.8488 & 0.8684 & 0.7510 & 0.8926 & 0.9035 & 0.8724 & 0.8561 \\
Llama-3.1-8B-Instruct-singprobe & \textbf{0.8586} & 0.8752 & 0.7788 & 0.9461 & 0.9218 & 0.8748 & 0.8759 \\
Qwen3-8B-singprobe & 0.8569 & 0.8680 & 0.7755 & 0.9307 & 0.9134 & 0.8678 & 0.8687 \\
Qwen3.5-9B-singprobe & 0.8473 & 0.8844 & 0.7543 & 0.9284 & 0.9342 & 0.8828 & 0.8719 \\
gpt-oss-20b-singprobe & 0.8502 & 0.8704 & 0.7855 & 0.9219 & 0.9124 & 0.8703 & 0.8685 \\
gemma-4-26B-A4B-it-singprobe & 0.8323 & 0.8437 & 0.7402 & 0.8952 & 0.8924 & 0.8598 & 0.8439 \\
Qwen3.5-27B-singprobe & 0.8560 & 0.8838 & 0.7607 & 0.9280 & 0.9313 & 0.8848 & 0.8741 \\
Qwen3.6-27B-singprobe & 0.8478 & 0.8667 & 0.7948 & 0.9330 & 0.9237 & 0.8648 & 0.8718 \\
Qwen3.8-27B-singprobe & 0.8482 & 0.8654 & 0.7818 & 0.9347 & 0.9185 & 0.8612 & 0.8683 \\
gemma-4-31B-it-singprobe & 0.8490 & 0.8721 & 0.7545 & 0.9429 & 0.9250 & 0.8765 & 0.8700 \\
Qwen3.5-35B-A3B-singprobe & 0.8487 & 0.8782 & 0.7844 & \textbf{0.9471} & 0.9188 & 0.8735 & 0.8751 \\
Qwen3.6-35B-A3B-singprobe & 0.8483 & 0.8821 & 0.7498 & 0.9238 & 0.9297 & 0.8856 & 0.8699 \\
gpt-oss-120b-singprobe & 0.8427 & 0.8770 & 0.7534 & 0.9291 & 0.9231 & 0.8940 & 0.8699 \\
Qwen3.5-122B-A10B-singprobe & 0.8513 & 0.8774 & 0.7797 & 0.9373 & 0.9016 & 0.8711 & 0.8697 \\
Ling-3.0-flash-singprobe & 0.8481 & 0.8795 & 0.7703 & 0.9012 & 0.9231 & 0.8825 & 0.8674 \\
Step-3.7-Flash-singprobe & 0.8493 & 0.7646 & 0.7723 & 0.9208 & 0.9240 & 0.8698 & 0.8502 \\
MiniMax-M2.7-singprobe & 0.8482 & 0.8735 & 0.7856 & 0.9310 & 0.9240 & 0.8764 & 0.8731 \\
DeepSeek-V4-Flash-0731-singprobe & 0.8536 & \textbf{0.8950} & 0.7730 & 0.9265 & \textbf{0.9571} & 0.8816 & 0.8811 \\
Hy3-singprobe & 0.8507 & 0.8868 & 0.7681 & 0.9181 & 0.9298 & 0.8790 & 0.8721 \\
Qwen3.5-397B-A17B-singprobe & 0.8505 & 0.8709 & \textbf{0.7962} & 0.9421 & 0.9195 & 0.8710 & 0.8750 \\
GLM-5.2-singprobe & 0.8494 & 0.8886 & 0.7278 & 0.9122 & 0.9411 & 0.8874 & 0.8677 \\
GLM-5.3-singprobe & 0.8455 & 0.8883 & 0.7611 & 0.9223 & 0.9457 & \textbf{0.9033} & 0.8777 \\
\bottomrule[1.2pt]
\end{tabular}
}
\end{table}

\subsection{Response Safety Evaluation}
\partitle{Response-level Classification}
Table~\ref{tab:cross_model_response} gives the response-unsafety F1 results. Qwen3.5-27B-singprobe obtains the highest macro F1 (0.8729), closely followed by Ling-3.0-flash-singprobe (0.8728). The narrow spread among many models---especially those above 4B parameters---indicates that the lightweight probe extracts a consistent response-safety signal across base-model families. No single base model dominates every dataset, which also highlights the importance of evaluating across heterogeneous safety distributions.

\begin{table}[t]
\centering
\scriptsize
\caption{Response-safety classification F1 by benchmark. Macro is the unweighted mean over the eight datasets.}
\vspace{0.5mm}
\label{tab:cross_model_response}
\resizebox{0.99\textwidth}{!}{%
\renewcommand{\arraystretch}{1.10}
\setlength{\tabcolsep}{3.8pt}
\begin{tabular}{lcccccccc>{\columncolor{green!8}}c}
\toprule[1.2pt]
\textbf{Model} & \textbf{BeaverTails-30k} & \textbf{PKU-SafeRLHF} & \textbf{AEGIS-2.0} & \textbf{WildGuard} & \textbf{XGuard} & \textbf{HarmBench} & \textbf{XSTest} & \textbf{ExpGuard} & \textbf{Avg.} \\
\midrule[0.8pt]
Qwen3-0.6B-singprobe & 0.8253 & 0.9167 & 0.7871 & 0.7589 & 0.7892 & 0.8801 & 0.8447 & 0.8831 & 0.8356 \\
Qwen3.5-0.8B-singprobe & 0.8229 & 0.9134 & 0.7933 & 0.7548 & 0.7953 & 0.8858 & 0.8364 & 0.8654 & 0.8334 \\
Llama-3.2-1B-Instruct-singprobe & 0.8356 & 0.9162 & 0.8227 & 0.7583 & 0.7955 & 0.8696 & 0.8810 & 0.8959 & 0.8468 \\
Qwen3.5-2B-singprobe & 0.8310 & 0.9225 & 0.8005 & 0.7616 & 0.8288 & 0.8784 & 0.8929 & 0.8974 & 0.8516 \\
gemma-4-E4B-it-singprobe & 0.8423 & 0.9301 & 0.8214 & 0.8000 & 0.8368 & 0.8872 & 0.9048 & 0.9219 & 0.8681 \\
Qwen3-4B-Instruct-2507-singprobe & 0.8423 & 0.9275 & 0.8030 & 0.7927 & 0.8201 & 0.8806 & 0.8862 & 0.9210 & 0.8592 \\
Qwen3.5-4B-singprobe & 0.8427 & 0.9286 & 0.8067 & 0.8052 & 0.8487 & 0.8810 & 0.8941 & 0.9099 & 0.8646 \\
Ling-3.0-tiny-singprobe & 0.8537 & 0.9120 & 0.8211 & 0.7683 & 0.8122 & 0.8645 & 0.8588 & 0.9159 & 0.8508 \\
Llama-3.1-8B-Instruct-singprobe & 0.8447 & 0.9229 & \textbf{0.8381} & 0.7987 & 0.8242 & 0.8777 & 0.8706 & 0.9224 & 0.8624 \\
Qwen3-8B-singprobe & 0.8511 & 0.9251 & 0.8161 & 0.7993 & 0.8353 & 0.8858 & 0.8686 & 0.9118 & 0.8616 \\
Qwen3.5-9B-singprobe & 0.8384 & 0.9327 & 0.8160 & 0.8027 & 0.8402 & 0.8973 & 0.9202 & 0.9088 & 0.8695 \\
gpt-oss-20b-singprobe & 0.8533 & 0.9225 & 0.8149 & 0.7890 & 0.8588 & 0.8763 & 0.8929 & 0.9075 & 0.8644 \\
gemma-4-26B-A4B-it-singprobe & 0.8432 & 0.9284 & 0.8071 & 0.8000 & 0.8348 & 0.8869 & 0.8889 & 0.9280 & 0.8647 \\
Qwen3.5-27B-singprobe & 0.8460 & 0.9317 & 0.8287 & 0.8013 & 0.8412 & 0.8947 & 0.9157 & 0.9243 & \textbf{0.8729} \\
Qwen3.6-27B-singprobe & 0.8502 & 0.9290 & 0.8291 & 0.7832 & 0.8512 & 0.8811 & 0.9048 & 0.9247 & 0.8692 \\
Qwen3.8-27B-singprobe & 0.8466 & 0.9250 & 0.8377 & 0.7974 & 0.8300 & 0.8947 & 0.8941 & 0.9153 & 0.8676 \\
gemma-4-31B-it-singprobe & 0.8430 & 0.9339 & 0.8079 & 0.7888 & 0.8353 & 0.8902 & 0.8957 & \textbf{0.9293} & 0.8655 \\
Qwen3.5-35B-A3B-singprobe & 0.8336 & 0.9377 & 0.8132 & 0.8000 & 0.8349 & 0.8885 & 0.9136 & 0.9163 & 0.8672 \\
Qwen3.6-35B-A3B-singprobe & 0.8477 & 0.9235 & 0.8286 & 0.7955 & 0.8290 & 0.8872 & 0.8941 & 0.9205 & 0.8658 \\
gpt-oss-120b-singprobe & 0.8443 & 0.9250 & 0.8139 & 0.7928 & 0.8455 & 0.8845 & 0.9102 & 0.9052 & 0.8652 \\
Qwen3.5-122B-A10B-singprobe & 0.8357 & 0.9345 & 0.8210 & 0.8027 & 0.8503 & 0.8750 & 0.9146 & 0.9102 & 0.8680 \\
Ling-3.0-flash-singprobe & 0.8433 & 0.9301 & 0.8333 & 0.8000 & \textbf{0.8588} & 0.8995 & 0.8929 & 0.9247 & 0.8728 \\
Step-3.7-Flash-singprobe & 0.8528 & 0.9224 & 0.8234 & 0.7968 & 0.7888 & 0.8800 & 0.8686 & 0.9115 & 0.8555 \\
MiniMax-M2.7-singprobe & 0.8333 & \textbf{0.9400} & 0.8153 & 0.8020 & 0.8468 & 0.8868 & \textbf{0.9268} & 0.9069 & 0.8698 \\
DeepSeek-V4-Flash-0731-singprobe & 0.8328 & 0.9392 & 0.8122 & 0.8088 & 0.8502 & 0.8924 & 0.9136 & 0.9089 & 0.8697 \\
Hy3-singprobe & \textbf{0.8538} & 0.9154 & 0.8289 & 0.7949 & 0.8459 & 0.8829 & 0.8721 & 0.9219 & 0.8645 \\
Qwen3.5-397B-A17B-singprobe & 0.8386 & 0.9344 & 0.8146 & 0.7960 & 0.8545 & 0.8899 & 0.9202 & 0.9085 & 0.8696 \\
GLM-5.2-singprobe & 0.8471 & 0.9290 & 0.8232 & 0.8000 & 0.8588 & 0.8985 & 0.8772 & 0.9220 & 0.8695 \\
GLM-5.3-singprobe & 0.8355 & 0.9384 & 0.8074 & \textbf{0.8115} & 0.8344 & \textbf{0.9012} & 0.9157 & 0.9274 & 0.8715 \\
\bottomrule[1.2pt]
\end{tabular}
}
\end{table}

\partitle{Streaming Safety Detection}
Table~\ref{tab:cross_model_streaming} reports response-level and token-level streaming AUC in the same format as the main-text streaming evaluation. Response-level discrimination is uniformly strong, with every model reaching at least 0.9740 mean AUC. Qwen3.5-397B-A17B-singprobe gives the highest mean response-level AUC (0.9905). For the more demanding token-localization metric, DeepSeek-V4-Flash-0731-singprobe performs best (0.9508), followed by Hy3-singprobe (0.9498), Qwen3-8B-singprobe (0.9484), MiniMax-M2.7-singprobe (0.9483), and Ling-3.0-flash-singprobe (0.9481).

\begin{table}[t]
\centering
\scriptsize
\caption{Cross-model streaming safety evaluation in terms of response-level AUC (R-AUC) and token-level AUC (T-AUC).}
\vspace{0.5mm}
\label{tab:cross_model_streaming}
\resizebox{0.95\textwidth}{!}{%
\renewcommand{\arraystretch}{1.10}
\setlength{\tabcolsep}{4.2pt}
\begin{tabular}{lcccccc>{\columncolor{green!8}}c>{\columncolor{green!8}}c}
\toprule[1.2pt]
\multirow{2}{*}{\textbf{Model}}
    & \multicolumn{2}{c}{\textbf{SingStreamBench}}
& \multicolumn{2}{c}{\textbf{SingStreamBench-Full}}
& \multicolumn{2}{c}{\textbf{FineHarm}}
& \multicolumn{2}{>{\columncolor{green!8}}c}{\textbf{Avg.}} \\
\cmidrule(lr){2-3}
\cmidrule(lr){4-5}
\cmidrule(lr){6-7}
\cmidrule(lr){8-9}
& \textbf{R-AUC} & \textbf{T-AUC}
& \textbf{R-AUC} & \textbf{T-AUC}
& \textbf{R-AUC} & \textbf{T-AUC}
& \textbf{R-AUC} & \textbf{T-AUC} \\
\midrule[0.8pt]
Qwen3-0.6B-singprobe & 0.9857 & 0.9472 & 0.9667 & 0.9251 & 0.9933 & 0.9508 & 0.9819 & 0.9410 \\
Qwen3.5-0.8B-singprobe & 0.9810 & 0.9207 & 0.9582 & 0.8829 & 0.9942 & 0.9581 & 0.9778 & 0.9206 \\
Llama-3.2-1B-Instruct-singprobe & 0.9838 & 0.9283 & 0.9680 & 0.8915 & 0.9932 & 0.9528 & 0.9817 & 0.9242 \\
Qwen3.5-2B-singprobe & 0.9797 & 0.9151 & 0.9722 & 0.9007 & 0.9954 & 0.9571 & 0.9824 & 0.9243 \\
gemma-4-E4B-it-singprobe & 0.9884 & 0.9152 & 0.9776 & 0.9085 & 0.9950 & 0.9552 & 0.9870 & 0.9263 \\
Qwen3-4B-Instruct-2507-singprobe & 0.9918 & 0.9498 & 0.9768 & 0.9332 & 0.9950 & 0.9575 & 0.9879 & 0.9468 \\
Qwen3.5-4B-singprobe & 0.9873 & 0.9206 & 0.9780 & 0.9064 & 0.9956 & 0.9588 & 0.9870 & 0.9286 \\
Ling-3.0-tiny-singprobe & 0.9935 & 0.9529 & 0.9775 & 0.9340 & 0.9954 & 0.9568 & 0.9888 & 0.9479 \\
Llama-3.1-8B-Instruct-singprobe & 0.9900 & 0.9254 & 0.9717 & 0.9034 & 0.9957 & 0.9585 & 0.9858 & 0.9291 \\
Qwen3-8B-singprobe & 0.9916 & 0.9564 & 0.9770 & 0.9355 & 0.9947 & 0.9533 & 0.9878 & 0.9484 \\
Qwen3.5-9B-singprobe & 0.9898 & 0.9328 & 0.9761 & 0.9084 & 0.9962 & \textbf{0.9619} & 0.9874 & 0.9344 \\
gpt-oss-20b-singprobe & 0.9824 & 0.9477 & 0.9573 & 0.9027 & 0.9947 & 0.9509 & 0.9781 & 0.9338 \\
gemma-4-26B-A4B-it-singprobe & 0.9703 & 0.9139 & 0.9560 & 0.8808 & 0.9959 & 0.9580 & 0.9740 & 0.9175 \\
Qwen3.5-27B-singprobe & 0.9902 & 0.9287 & 0.9769 & 0.8979 & 0.9965 & 0.9602 & 0.9879 & 0.9289 \\
Qwen3.6-27B-singprobe & 0.9876 & 0.9265 & 0.9753 & 0.8945 & 0.9965 & 0.9598 & 0.9864 & 0.9270 \\
Qwen3.8-27B-singprobe & 0.9857 & 0.9182 & 0.9820 & 0.9123 & 0.9965 & 0.9608 & 0.9881 & 0.9305 \\
gemma-4-31B-it-singprobe & 0.9864 & 0.9205 & 0.9675 & 0.8809 & 0.9953 & 0.9539 & 0.9830 & 0.9184 \\
Qwen3.5-35B-A3B-singprobe & 0.9935 & 0.9402 & 0.9810 & 0.9132 & 0.9960 & 0.9587 & 0.9902 & 0.9374 \\
Qwen3.6-35B-A3B-singprobe & 0.9927 & 0.9414 & 0.9778 & 0.9072 & 0.9963 & 0.9591 & 0.9889 & 0.9359 \\
gpt-oss-120b-singprobe & \textbf{0.9953} & 0.9568 & 0.9736 & 0.9274 & 0.9955 & 0.9520 & 0.9881 & 0.9454 \\
Qwen3.5-122B-A10B-singprobe & 0.9908 & 0.9258 & 0.9814 & 0.9100 & 0.9963 & 0.9553 & 0.9895 & 0.9304 \\
Ling-3.0-flash-singprobe & 0.9922 & \textbf{0.9599} & 0.9778 & 0.9294 & 0.9961 & 0.9549 & 0.9887 & 0.9481 \\
Step-3.7-Flash-singprobe & 0.9878 & 0.9413 & 0.9740 & 0.9122 & 0.9956 & 0.9350 & 0.9858 & 0.9295 \\
MiniMax-M2.7-singprobe & 0.9925 & 0.9505 & 0.9813 & 0.9372 & 0.9968 & 0.9572 & 0.9902 & 0.9483 \\
DeepSeek-V4-Flash-0731-singprobe & 0.9915 & 0.9577 & 0.9777 & 0.9411 & 0.9955 & 0.9536 & 0.9882 & \textbf{0.9508} \\
Hy3-singprobe & 0.9917 & 0.9499 & \textbf{0.9821} & \textbf{0.9429} & 0.9958 & 0.9567 & 0.9899 & 0.9498 \\
Qwen3.5-397B-A17B-singprobe & 0.9926 & 0.9332 & 0.9819 & 0.9076 & \textbf{0.9968} & 0.9608 & \textbf{0.9905} & 0.9339 \\
GLM-5.2-singprobe & 0.9873 & 0.9333 & 0.9723 & 0.9028 & 0.9960 & 0.9562 & 0.9852 & 0.9308 \\
GLM-5.3-singprobe & 0.9902 & 0.9311 & 0.9768 & 0.9062 & 0.9961 & 0.9592 & 0.9877 & 0.9322 \\
\bottomrule[1.2pt]
\end{tabular}
}
\end{table}

\partitle{False Positives on Benign Responses}
Table~\ref{tab:cross_model_fpr} reports false-positive rates on benign model generations. False positives are rare for nearly all probes and are concentrated in Dolly-15k. DeepSeek-V4-Flash-0731-singprobe has the lowest pooled FPR (0.01\%); Qwen3.6-27B-singprobe and Hy3-singprobe follow at 0.02\%.

\begin{table}[t]
\centering
\scriptsize
\caption{Response-unsafety false-positive rate on benign instruction datasets. Pooled is computed over all examples (lower is better).}
\vspace{0.5mm}
\label{tab:cross_model_fpr}
\resizebox{0.9\textwidth}{!}{%
\renewcommand{\arraystretch}{1.10}
\setlength{\tabcolsep}{4.2pt}
\begin{tabular}{lccccc>{\columncolor{green!8}}c}
\toprule[1.2pt]
\textbf{Model} & \textbf{MATH} & \textbf{GSM8K} & \textbf{Dolly-15k} & \textbf{MBPP} & \textbf{OpenBookQA} & \textbf{Avg.} \\
\midrule[0.8pt]
Qwen3-0.6B-singprobe & 0.00\% & 0.00\% & 0.50\% & 0.00\% & 0.02\% & 0.21\% \\
Qwen3.5-0.8B-singprobe & 0.00\% & 0.08\% & 0.55\% & 0.00\% & 0.00\% & 0.23\% \\
Llama-3.2-1B-Instruct-singprobe & 0.00\% & 0.00\% & 0.20\% & 0.00\% & 0.00\% & 0.08\% \\
Qwen3.5-2B-singprobe & 0.00\% & 0.00\% & 0.17\% & 0.00\% & 0.05\% & 0.08\% \\
gemma-4-E4B-it-singprobe & 0.00\% & 0.00\% & 0.09\% & 0.00\% & 0.00\% & 0.04\% \\
Qwen3-4B-Instruct-2507-singprobe & 0.00\% & 0.00\% & 0.15\% & 0.00\% & 0.00\% & 0.06\% \\
Qwen3.5-4B-singprobe & 0.00\% & 0.00\% & 0.09\% & 0.00\% & 0.00\% & 0.04\% \\
Ling-3.0-tiny-singprobe & 0.00\% & 0.00\% & 0.14\% & 0.00\% & 0.05\% & 0.07\% \\
Llama-3.1-8B-Instruct-singprobe & 0.00\% & 0.00\% & 0.07\% & 0.00\% & 0.00\% & 0.03\% \\
Qwen3-8B-singprobe & 0.00\% & 0.00\% & 0.08\% & 0.00\% & 0.00\% & 0.03\% \\
Qwen3.5-9B-singprobe & 0.00\% & 0.00\% & 0.09\% & 0.00\% & 0.00\% & 0.04\% \\
gpt-oss-20b-singprobe & 0.00\% & 0.00\% & 0.11\% & 0.00\% & 0.00\% & 0.05\% \\
gemma-4-26B-A4B-it-singprobe & 0.00\% & 0.00\% & 0.09\% & 0.00\% & 0.03\% & 0.04\% \\
Qwen3.5-27B-singprobe & 0.00\% & 0.00\% & 0.06\% & 0.00\% & 0.00\% & 0.03\% \\
Qwen3.6-27B-singprobe & 0.00\% & 0.00\% & 0.05\% & 0.00\% & 0.00\% & 0.02\% \\
Qwen3.8-27B-singprobe & 0.00\% & 0.00\% & 0.06\% & 0.00\% & 0.00\% & 0.03\% \\
gemma-4-31B-it-singprobe & 0.00\% & 0.00\% & 0.05\% & 0.00\% & 0.02\% & 0.03\% \\
Qwen3.5-35B-A3B-singprobe & 0.00\% & 0.00\% & 0.07\% & 0.00\% & 0.00\% & 0.03\% \\
Qwen3.6-35B-A3B-singprobe & 0.00\% & 0.00\% & 0.07\% & 0.00\% & 0.00\% & 0.03\% \\
gpt-oss-120b-singprobe & 0.00\% & 0.00\% & 0.14\% & 0.00\% & 0.00\% & 0.06\% \\
Qwen3.5-122B-A10B-singprobe & 0.00\% & 0.00\% & 0.06\% & 0.00\% & 0.00\% & 0.03\% \\
Ling-3.0-flash-singprobe & 0.00\% & 0.00\% & 0.07\% & 0.00\% & 0.00\% & 0.03\% \\
Step-3.7-Flash-singprobe & 0.01\% & 0.00\% & 0.12\% & 0.00\% & 0.02\% & 0.05\% \\
MiniMax-M2.7-singprobe & 0.00\% & 0.00\% & 0.06\% & 0.00\% & 0.02\% & 0.03\% \\
DeepSeek-V4-Flash-0731-singprobe & 0.00\% & 0.00\% & 0.03\% & 0.00\% & 0.00\% & \textbf{0.01\%} \\
Hy3-singprobe & 0.00\% & 0.00\% & 0.05\% & 0.00\% & 0.00\% & 0.02\% \\
Qwen3.5-397B-A17B-singprobe & 0.00\% & 0.00\% & 0.08\% & 0.00\% & 0.00\% & 0.03\% \\
GLM-5.2-singprobe & 0.00\% & 0.00\% & 0.05\% & 0.00\% & 0.00\% & 0.02\% \\
GLM-5.3-singprobe & 0.00\% & 0.00\% & 0.06\% & 0.00\% & 0.00\% & 0.03\% \\
\bottomrule[1.2pt]
\end{tabular}
}
\end{table}

\subsection{Response Hallucination Evaluation}
Table~\ref{tab:cross_model_hallu} reports hallucination-detection AUC. GLM-5.3-singprobe attains the highest displayed macro AUC (0.8170), followed by GLM-5.2-singprobe (0.8118), Qwen3.5-35B-A3B-singprobe and Qwen3.5-397B-A17B-singprobe (both 0.8117), and Qwen3.6-27B-singprobe (0.8094). FaithDial yields consistently high AUC, while FAVA is generally the most challenging benchmark.

\begin{table}[t!]
\centering
\scriptsize
\caption{Hallucination-detection AUC by benchmark. Macro is the unweighted mean over the six datasets.}
\vspace{0.5mm}
\label{tab:cross_model_hallu}
\resizebox{0.95\textwidth}{!}{%
\renewcommand{\arraystretch}{1.10}
\setlength{\tabcolsep}{3.8pt}
\begin{tabular}{lcccccc>{\columncolor{green!8}}c}
\toprule[1.2pt]
\textbf{Model} & \textbf{FactCHD} & \textbf{FaithDial} & \textbf{FAVA} & \textbf{RAGTruth} & \textbf{Shroom} & \textbf{WikiBio-GPT3} & \textbf{Avg.} \\
\midrule[0.8pt]
Qwen3-0.6B-singprobe & 0.7539 & 0.9532 & 0.5811 & 0.6432 & 0.6636 & 0.9602 & 0.7592 \\
Qwen3.5-0.8B-singprobe & 0.7620 & 0.9519 & 0.5859 & 0.6417 & 0.6487 & 0.8977 & 0.7480 \\
Llama-3.2-1B-Instruct-singprobe & 0.7418 & 0.9438 & 0.6224 & 0.6227 & 0.6505 & 0.9463 & 0.7546 \\
Qwen3.5-2B-singprobe & 0.7630 & 0.9502 & 0.5809 & 0.6342 & 0.7118 & 0.9451 & 0.7642 \\
gemma-4-E4B-it-singprobe & 0.7777 & 0.9520 & 0.6091 & 0.7169 & 0.7336 & 0.8761 & 0.7776 \\
Qwen3-4B-Instruct-2507-singprobe & 0.7974 & 0.9490 & 0.6479 & 0.6917 & 0.7227 & 0.9440 & 0.7921 \\
Qwen3.5-4B-singprobe & 0.7985 & 0.9517 & 0.5978 & 0.7266 & 0.7338 & 0.9047 & 0.7855 \\
Ling-3.0-tiny-singprobe & 0.7819 & 0.9529 & 0.5802 & 0.6980 & 0.6946 & 0.9515 & 0.7765 \\
Llama-3.1-8B-Instruct-singprobe & 0.7937 & 0.9449 & 0.6088 & 0.6705 & 0.7355 & 0.8687 & 0.7704 \\
Qwen3-8B-singprobe & 0.7933 & 0.9478 & 0.6173 & 0.6869 & 0.7420 & 0.9222 & 0.7849 \\
Qwen3.5-9B-singprobe & 0.8076 & 0.9478 & 0.6499 & 0.7349 & 0.7384 & 0.8940 & 0.7954 \\
gpt-oss-20b-singprobe & 0.7974 & 0.9457 & 0.5926 & 0.6124 & 0.7345 & 0.8660 & 0.7581 \\
gemma-4-26B-A4B-it-singprobe & 0.8032 & 0.9460 & 0.6258 & 0.7632 & 0.7638 & 0.8695 & 0.7952 \\
Qwen3.5-27B-singprobe & 0.8244 & 0.9487 & 0.5988 & 0.7093 & 0.7659 & 0.9331 & 0.7967 \\
Qwen3.6-27B-singprobe & 0.8199 & 0.9547 & 0.6271 & 0.7706 & 0.7625 & 0.9215 & 0.8094 \\
Qwen3.8-27B-singprobe & 0.8168 & 0.9516 & 0.6155 & 0.7049 & 0.7581 & \textbf{0.9738} & 0.8035 \\
gemma-4-31B-it-singprobe & 0.8170 & 0.9468 & 0.6111 & 0.7551 & 0.7763 & 0.9261 & 0.8054 \\
Qwen3.5-35B-A3B-singprobe & 0.8262 & 0.9577 & 0.6111 & 0.7335 & 0.7747 & 0.9672 & 0.8117 \\
Qwen3.6-35B-A3B-singprobe & 0.8253 & \textbf{0.9583} & 0.5961 & 0.7296 & 0.7639 & 0.9403 & 0.8023 \\
gpt-oss-120b-singprobe & 0.7979 & 0.9426 & 0.6159 & 0.6770 & 0.7407 & 0.9150 & 0.7815 \\
Qwen3.5-122B-A10B-singprobe & 0.8241 & 0.9496 & 0.5581 & 0.6588 & 0.7628 & 0.9299 & 0.7806 \\
Ling-3.0-flash-singprobe & 0.8308 & 0.9498 & 0.5996 & 0.7282 & 0.7314 & \textbf{0.9674} & 0.8012 \\
Step-3.7-Flash-singprobe & 0.8204 & 0.9430 & 0.6378 & 0.7046 & 0.6858 & 0.9510 & 0.7904 \\
MiniMax-M2.7-singprobe & 0.8112 & 0.9503 & 0.5708 & 0.6498 & 0.7511 & 0.9333 & 0.7777 \\
DeepSeek-V4-Flash-0731-singprobe & 0.8173 & 0.9427 & 0.6336 & 0.6891 & 0.7537 & 0.9327 & 0.7949 \\
Hy3-singprobe & 0.8313 & 0.9488 & 0.6401 & 0.7088 & 0.7667 & 0.9447 & 0.8067 \\
Qwen3.5-397B-A17B-singprobe & \textbf{0.8423} & 0.9554 & \textbf{0.6561} & 0.6770 & 0.7756 & 0.9639 & 0.8117 \\
GLM-5.2-singprobe & 0.8353 & 0.9507 & 0.5900 & \textbf{0.7736} & \textbf{0.7799} & 0.9410 & 0.8118 \\
GLM-5.3-singprobe & 0.8366 & 0.9475 & 0.6183 & 0.7634 & 0.7794 & 0.9565 & 0.8170 \\
\bottomrule[1.2pt]
\end{tabular}
}
\end{table}

\end{document}